\documentclass[prd,aps,twocolumn,nofootinbib,superscriptaddress,floatfix,preprintnumbers]{revtex4-1}

\usepackage{mathrsfs, amsmath, amssymb}
\usepackage{diagbox, slashed, makecell}
\usepackage{url}
\usepackage{color}
\usepackage{xcolor}
\usepackage{ulem}
\usepackage{graphicx}
\usepackage{appendix}
\usepackage{verbatim}
\usepackage{multirow}
\usepackage{ucs}
\usepackage{enumitem}
\usepackage{upgreek}
\usepackage[colorlinks=true,citecolor=blue, linkcolor=blue, urlcolor=blue]{hyperref}

\allowdisplaybreaks

\graphicspath{{figures/}}

\begin{document}



\title{Loop-Level Lepton Flavor Violation and Diphoton Signals in the Minimal Left-Right Symmetric Model}

\author{Shufang Qiang}
\email{sfqiang@seu.edu.cn}
\affiliation{School of Physics, Southeast University, Nanjing 211189, P. R. China}
\author{Peiwen Wu}
\email{Corresponding author, pwwu@seu.edu.cn}
\affiliation{School of Physics, Southeast University, Nanjing 211189, P. R. China}
\author{Yongchao Zhang}
\email{Corresponding author, zhangyongchao@seu.edu.cn}
\affiliation{School of Physics, Southeast University, Nanjing 211189, P. R. China}
\affiliation{Center for High Energy Physics, Peking University, Beijing 100871, China}

\date{\today}

\begin{abstract}
The left-right symmetric model (LRSM) could not only restore parity of the weak interaction, but also provide natural explanations of the tiny active neutrino masses via the seesaw mechanisms. The $SU(2)_R$-breaking scalar $H_3$ can induce lepton flavor violating (LFV) effects in the minimal version of LRSM at the 1-loop order, originating from the mixing of heavy right-handed neutrinos (RHNs). If $H_3$ is light, say below the GeV scale, it will lead to rich signals, e.g. the LFV muon and tauon decays $\ell_\beta \to \ell_\alpha + X$ ($X$ being either visible or invisible final states) and the anomalous supernova signatures. Combined with the diphoton coupling of $H_3$, and recasting the existing constraints onto the light $H_3$ scenario, the right-handed scale $v_R$ is excluded up to $2\times10^9$ GeV. In the future, the $v_R$ scale can be probed up to $5\times10^9$ GeV in high-precision muon experiments, if the Yukwa couplings for RHN masses are of order one and the RHN mixing is maximal, and further up to $6\times10^{11}$ GeV by supernova observations, reaching the non-resonant leptogenesis scale in the LRSM.
\end{abstract}

\maketitle



\section{Introduction}

The precision of mixing of active neutrinos has been improved significantly in the last decades~\cite{ParticleDataGroup:2024cfk}, in particular in light of the very recent JUNO data~\cite{JUNO:2025gmd} (the matter effects are about 4\% on the oscillation parameters~\cite{Khan:2019doq}).
The tiny neutrino masses can be explained naturally and economically by the seesaw mechanism. The left-right symmetric model (LRSM)~\cite{Pati:1974yy,Mohapatra:1974gc,Senjanovic:1975rk}, based on the gauge group of $SU(3)_C \times SU(2)_L \times SU(2)_R \times U(1)_{B-L}$, was originally proposed to understand  parity violation in the weak interaction~\cite{Lee:1956qn,Wu:1957my}, and offers a well-motivated framework for the type-I~\cite{Minkowski:1977sc,Mohapatra:1979ia,Yanagida:1979as,Gell-Mann:1979vob,Glashow:1979nm} and type-II~\cite{Mohapatra:1980yp,Magg:1980ut,Schechter:1980gr,Cheng:1980qt,Lazarides:1980nt} seesaw mechanisms. 

In the conventional minimal version of LRSM, the $SU(2)_R$ gauge symmetry and parity is broken spontaneously by a triplet scalar $\Delta_R$. The neutral CP-even component $H_3$ of $\Delta_R$ plays a central role, not only in the breaking of the $SU(2)_R$ gauge symmetry, but also for generation of the masses of the heavy scalars, right-handed neutrinos (RHNs) and the $W_R$ and $Z_R$ bosons. The relevant rich phenomenologies have been investigated extensively in the literature~\cite{Deshpande:1990ip,Zhang:2007da,Blanke:2011ry,Maiezza:2015lza,Dev:2016dja,Maiezza:2016bzp,Nemevsek:2016enw,Maiezza:2016ybz,BhupalDev:2016nfr,Dev:2017dui,BhupalDev:2018vpr,BhupalDev:2018xya,Chauhan:2018uuy,Chauhan:2019fji,Brdar:2019fur,Li:2020eun,Borah:2022wdy,Borboruah:2022eex,Kriewald:2024cgr,Wang:2024wcs,Dev:2025fcv,Searle:2025cnj}. A unique feature of $H_3$ lies in the fact that it does not couple directly to the SM particles, which has far reaching phenomenological implications. Unlike other particles in LRSM, the direct constraints on $H_3$ are rather weak. In certain regions of parameter space, $H_3$ could be very light with respect to the $SU(2)_R$-breaking $v_R$ scale~\cite{BhupalDev:2016nfr,Dev:2017dui,Nemevsek:2016enw}.
For $v_R \gtrsim 10^{15}$ GeV $H_3$ can even play the role of decaying dark matter if its mass is below the MeV scale~\cite{Dev:2025fcv}.

\begin{figure}
    \centering
    \includegraphics[width= 0.325 \linewidth]{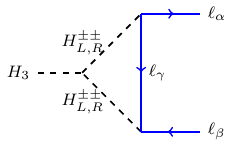} 
    \includegraphics[width= 0.325 \linewidth]{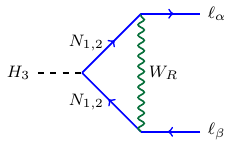}
    \includegraphics[width= 0.325 \linewidth]{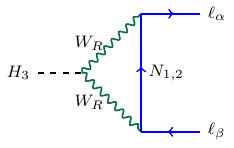}
    \caption{The 1-loop Feynman diagrams for the LFV couplings of $H_3$ in the minimal LRSM. }
    \label{fig:diagram}
    \vspace{-10pt}
\end{figure}

For a light $H_3$ with mass $m_{H_3}$ below the GeV scale, its couplings to the standard model (SM) particles can arise from mixing with the SM Higgs and the heavy bidoublet scalar $H_1$. The resultant rich signals such as $B \to K + H_3$ have been studied in great details in Refs.~\cite{BhupalDev:2016nfr,Dev:2017dui}. $H_3$ can also couple to the SM particles at the 1-loop order, which is induced by the heavy particles in the LRSM. A typical example is the coupling of $H_3$ to two photons. In this paper, we focus more on the radiative couplings of $H_3$ to charged leptons at the 1-loop order, induced by the doubly-charged scalars, $W_R$ boson and RHNs $N_i$~\cite{Nemevsek:2016enw}. The Feynman diagrams are shown in Fig.~\ref{fig:diagram}. Decoupling the left-handed doubly-charged scalar, the charged lepton flavor violating (LFV) couplings of $H_3$ intrinsically originate from the mixing of RHNs $N_i$ via the Yukawa coupling matrix $f_R$ of $H_3$ with $N_i$. This provides an well-motivated example to build up connects between charged LFV and oscillations of active neutrinos.

The LFV couplings of light $H_3$ can lead to very rich interesting signals, which depend on the mass $m_{H_3}$ and the $v_R$ scale, as all the heavy particles masses in the loops in Fig.~\ref{fig:diagram} are at the $v_R$ scale.
In some sense, this is  very similar to the dependence of the couplings of axion-like particles (ALPs) on the decay constant $f_a$. It turns out that the most stringent LFV constraints are from the rare muon decays $\mu \to e + {\rm inv}$~\cite{TWIST:2014ymv, PIENU:2020loi, Jodidio:1986mz}, $\tau \to \ell + X$ (with $\ell = e,\,\mu$, and $X$ invisible or visible)~\cite{Belle:2025bpu,Ema:2025bww} and the supernova limits~\cite{Li:2025beu, Huang:2025rmy,Huang:2025xvo}. With Yukawa couplings of order one and the maximal mixing of RHNs, the $SU(2)_R$-breaking $v_R$ scale is excluded up to $4\times10^8$ GeV.
The supernova constraints on the diphoton coupling of $H_3$ are more stringent~\cite{Muller:2023vjm,Muller:2023pip,Fiorillo:2025yzf}, excluding $v_R$ up to $2\times10^9$ GeV.
These limits are orders of magnitude higher than the direct LHC limits from the direct searches of $W_R$ boson~\cite{ATLAS:2023cjo,CMS:2021dzb}, the constraints from the LFV effects induced by the doubly-charged scalars~\cite{Cirigliano:2004mv,Cirigliano:2004tc,Tello:2010am,Nemevsek:2011aa,Das:2012ii,Barry:2013xxa,Lee:2013htl,Deppisch:2014zta,Bambhaniya:2015ipg,Borah:2016iqd,Bonilla:2016fqd,BhupalDev:2018tox,Alves:2022yav} and the $W_R$ boson and $N_i$~\cite{Tello:2010am,Alonso:2012ji,Dinh:2012bp,Keung:1983uu,Cirigliano:2004mv,Akeroyd:2006bb,Aguilar-Saavedra:2012dga,Das:2012ii,Awasthi:2013ff,Barry:2013xxa,Lee:2013htl,Deppisch:2014zta,Bonilla:2016fqd,Lindner:2016bgg,Das:2017hmg}.
It is remarkable that the sensitivities of the $v_R$ scale can be improved up to $5\times10^9$ GeV in future high-precision muon experiments via the LFV coupling, and further higher up to $6\times10^{11}$ GeV by future supernova observations via the diphoton coupling, reaching the non-resonant leptogenesis scale in the LRSM~\cite{Frere:2008ct,ReFiorentin:2016rzn}. For small Yukawa couplings or small mixing angle of RHNs, the current LFV limits on the right-handed scale $v_R$ and future prospects will be weakened accordingly.

\section{LFV couplings of $H_3$}

In the minimal LRSM, the scalar sector consists of a Higgs bidoublet $\Phi$, a left-handed triplet $\Delta_L$ and a right-handed triplet $\Delta_R$:
\begin{equation}
\label{eq:scalar}
\Phi = \left(\begin{array}{cc}\phi^0_1 & \phi^+_2\\\phi^-_1 & \phi^0_2\end{array}\right) , \;
\Delta_{L,R} = \left(\begin{array}{cc}\Delta^+_{L,R}/\sqrt{2} & \Delta^{++}_{L,R} \\ \Delta^0_{L,R} & -\Delta^+_{L,R}/\sqrt{2}\end{array} \right) \,.
\end{equation}
The left-right gauge symmetry is broken by the vacuum expectation value (VEV) of the real component of $\Delta_R^0$ to the SM gauge group. The corresponding physical scalar field is $H_3$,
which is responsible for the mass generation of the heavy scalars, RHNs and the $W_R$ and $Z_R$ bosons in the LRSM. Even if the radiative corrections of all these heavy LRSM particles to its mass $m_{H_3}$ are taken into account, $H_3$ could be much lighter than the $v_R$ scale in some regions of the parameter space~\cite{BhupalDev:2016nfr,Dev:2017dui}. The constraint of electroweak precision data on the large mass splitting of $H_3$ with other components of $\Delta_R$ via the oblique parameters are also rather weak, suppressed by $v_{}^2/v_R^2$ (with $v_{}$ the electroweak scale)~\cite{Roitgrund:2020cge}.

In the limit of vanishing mixing with the SM Higgs and other neutral scalars in the LRSM, the scalar $H_3$ does not couple directly to charged leptons. Such couplings can be induced at the 1-loop order by the couplings of $H_3$ with the left- and right-handed doubly-charged scalars $H_{L,R}^{\pm\pm}$, RHNs $N_i$ and the $W_R$ boson, as seen in Fig.~\ref{fig:diagram}. In the first diagram, the couplings of $H_L^{\pm\pm}$ to charged leptons are from the term of $f_{L} \psi_{L}^{\sf T} C i \sigma_2 \Delta_L \psi_{L} +{\rm H.c.}$ (with $\psi$ the lepton doublet, $\sigma_2$ the second Pauli matrix and $C$ the charge conjugation operator), which is responsible for generation of active neutrino masses via the type-II seesaw. The rest diagrams in Fig.~\ref{fig:diagram} are all relevant to the Yukawa couplings of $H_3$ with the heavy RHNs:
\begin{eqnarray}
\label{eqn:Lyukawa}
\mathcal{L}_Y =
- f_{R} \psi_{R}^{\sf T} C i \sigma_2 \Delta_R \psi_{R} + \text{H.c.} \,,
\end{eqnarray}
where $f_R$ is the Yukawa coupling matrix and plays a central role in this paper. In particular, the mass matrix for the heavy RHNs is $M_N = f_R v_R$. In minimal LRSM the matrices $f_{L,\,R}$ can be different from each other. The left-handed triplet $\Delta_L$ can even decouple from the low-energy physics in the $D$-parity violating case~\cite{Chang:1983fu}. For simplicity, we neglect the contribution from $\Delta_L$ and consider in this paper the LFV effects only from the RHNs via the matrix $f_R$. For illustration purpose, we adopt the simplest case with only two RHNs $N_{1,2}$, with the Yukawa couplings $f_{1,2}$ and mixing angle $\theta$ without any CP violation. Then the LFV effects originate from the mixing of $N_{1,2}$.

Without loss of generality, the LFV couplings of $H_3$ with charged leptons can be written as
\begin{equation}
\label{eqn:LFV}
\mathcal{L}_{\rm LFV} = - H_3 \bar{\ell}_\alpha \left(c_{\alpha\beta}^{(L)}\, P_L + c_{\alpha\beta}^{(R)}\, P_R \right) \ell_\beta  ~+~ {\rm H.c.} \,,
\end{equation}
where $\alpha,\, \beta = e,\, \mu,\,\tau$, and $c_{\alpha\beta}^{(L)}$ and $c_{\alpha\beta}^{(R)}$ are dimensionless functions of the masses of $H_3$, charged leptons and the heavy particles $H_R^{\pm\pm}$, $W_R$ and $N_{1,2}$, given in Eq.~(\ref{eqn:cLcR})~\cite{Hahn:1998yk}. In some cases it is more convenient to use the combination of $g_{\alpha\beta} \equiv \sqrt{|c_{\alpha\beta}^{(L)}|^2 + |c_{\alpha\beta}^{(R)}|^2}$ for the following calculations.
For concreteness, we setup the parameters as follows. ({\it i}) The right-handed gauge coupling $g_R$ is set to be the same as the gauge coupling $g_L$ for $SU(2)_L$. ({\it ii}) We take the quartic coupling $\rho_2 = 0.25$, such that the right-handed doubly-charged mass $m_{H_R^{\pm\pm}} = \sqrt{4\rho_2} v_R = v_R$. ({\it iii}) The LFV couplings in Eq.~(\ref{eqn:LFV}) are rather sensitive to the Yukawa couplings $f_{1,2}$.
As a benchmark study, we adopt the values of $f_1 = 0.5$ and $f_2 = 1.0$.
({\it iv}) The RHN mixing is set to be $\theta = 45^\circ$ to maximize the LFV effect. The analysis of other benchmark scenarios can be found in a followup paper~\cite{Qiang:2026abc}.

It should be noted that the RHN mixing does not only leads to the LFV couplings in Eq.~(\ref{eqn:LFV}), but also generates the LFC couplings $g_\ell H_3 \ell^+ \ell^-$, the corresponding effective coupling $g_\ell$ is given in Eq.~(\ref{eqn:gl}). Another important decay channel for a light $H_3$ in the LRSM is the diphoton mode, i.e. $H_3 \to \gamma\gamma$, which is mediated by the singly- and doubly-charged scalars and the $W_R$ boson~\cite{BhupalDev:2016nfr,Dev:2017dui,Nemevsek:2016enw}.
For our benchmark point values of $f_{1,2}$, the dilepton decay branching ratios (BRs) are comparable to that of the diphoton mode, depending on the mass $m_{H_3}$ (cf. Fig.~\ref{fig:BR} in Appendix~\ref{app:width}).

\section{Laboratory LFV constraints}

\begin{figure*}
    \centering
    \includegraphics[width= 0.85\linewidth]{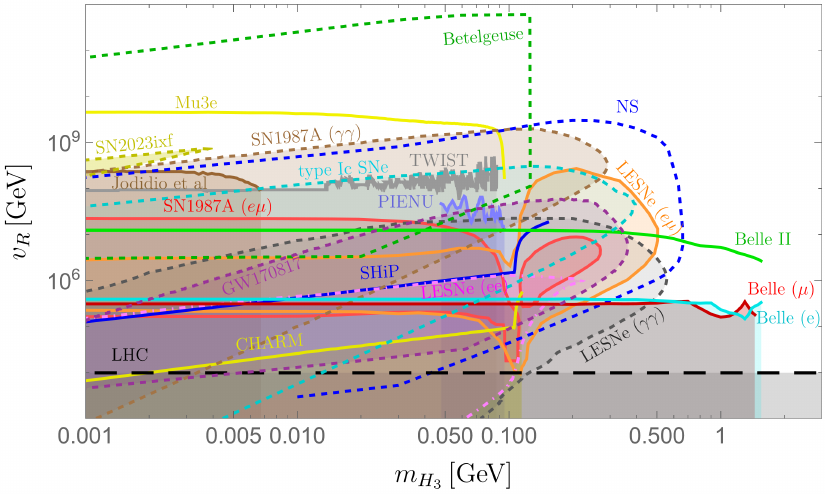}
    \caption{Constraints on $m_{H_3}$ and $v_R$ from the LFV, LFC and diphoton couplings of $H_3$ in the minimal LRSM, labeled as solid, dot-dashed and dashed lines, respectively.
    The shaded regions are for current limits, and the lines are for future prospects. The long dashed back line indicates the current LHC limits from direct searches of heavy $W_R$ boson~\cite{ATLAS:2023cjo,CMS:2021dzb}. See text and Tables~\ref{tab:LFV} and \ref{tab:gauge} for details.}
    \label{fig:LFV}
\end{figure*}

The LFV, LFC and diphoton channels of $H_3$ are all effectively suppressed by the $SU(2)$-breaking $v_R$ scale.
For $m_{H_3} \lesssim 1$ GeV, $H_3$ always behaves as a long-lived particle in the laboratory experiments (cf. Fig.~\ref{fig:BR} in Appendix~\ref{app:width}), and the leading current laboratory constraints are from the following processes, which are collected in Fig.~\ref{fig:LFV}~\cite{Dev:2017ftk,BhupalDev:2018vpr,Escribano:2020wua,Calibbi:2020jvd,Bauer:2021mvw,Cheung:2021mol,Davidson:2022jai,Han:2020dwo,Han:2022iig,Ma:2021jkp,Cui:2021dkr,Panci:2022wlc,Badziak:2024szg,Badziak:2025mkt,Wang:2025xyh,Greljo:2025ljr}. More details can be found in Table~\ref{tab:LFV} in Appendix~\ref{app:limit}.

{\it (i) Muon decays}. A light $H_3$ can be produced in the LFV process $\mu \to e + H_3$.
As $H_3$ is very long-lived, the signal is $\mu \to e + {\rm inv}$ at muon experiments. The current most stringent constraints in this channel are from the experiments TWIST~\cite{TWIST:2014ymv}, PIENU~\cite{PIENU:2020loi} and that by Jodiddo {\it et al}~\cite{Jodidio:1986mz} (see Refs.~\cite{Bolton:1988af,Calibbi:2020jvd,Bryman:1986wn} for weaker limits). ${\rm BR}(\mu \to e + {\rm inv})$ is excluded up to $2.6 \times 10^{-6}$, which leads to the limit of $v_R \gtrsim 5\times 10^{8}$ GeV.
The precision of $\mu \to e + {\rm inv}$ can be largely improved at future muon experiments. With a total number of $2.6\times 10^{15}$ muon events, the sensitivity of ${\rm BR} (\mu \to e + {\rm inv})$ can be improved up to ${\cal O} (10^{-8})$ at Mu3e~\cite{Perrevoort:2018ttp,Banerjee:2022nbr} (see Refs.~\cite{Banerjee:2022nbr,PIONEER:2022alm,Hill:2023dym,Calibbi:2020jvd,Jho:2022snj,Knapen:2023zgi} for weaker prospects). 
The corresponding prospect of $v_R$ scale can be improved by one order of magnitude up to $5\times 10^9$ GeV.

{\it (ii) Tauon decays}.
We can also have the rare LFV tauon decays $\tau \to \ell + H_3$ (with $\ell = e,\, \mu$).
The most stringent limits at $B$-factories are from the invisible decays $\tau \to \ell + {\rm inv}$. The scalar $H_3$ with a GeV-scale mass is not totally invisible, and would induce displaced vertex signatures. To be conservative, we require $H_3$ to decay outside the detectors by multiplying the factor of $\exp \{ - L_{\rm det}/\gamma_{H_3} \tau_{H_3} \}$,
with $\gamma_{H_3}$ is the Lorentz boost factor of $H_3$, and $L_{\rm det}$ the detector size. It turns out that the most constraining one is from the measurements of Belle, with ${\rm BR} (\tau \to e + {\rm inv}) <  (0.4-6.4) \times 10^{-4}$ and ${\rm BR} (\tau \to \mu + {\rm inv}) <  (0.2-3.5) \times 10^{-4}$.
The Belle detector is roughly 8 m in length and 8 m in diameter, which generates $L_{\rm det} \simeq 8.94$ m~\cite{Belle:2000cnh}.
This leads to the limits of $v_R \gtrsim 3\times10^5$ and $4\times10^5$ GeV in the $e$ and $\mu$ channels, respectively.
The constraints from ARGUS~\cite{ARGUS:1995bjh}, Belle II~\cite{Belle-II:2022heu} and Bryman {\it et al}~\cite{Bryman:2021ilc} are relatively weaker.
The sensitivity of $\tau \to \ell + {\rm inv}$ can be greatly improved at future $B$-factories. With an integrated luminosity of 50 ab\(^{-1}\), ${\rm BR}(\tau \to e + {\rm inv})$ can be measured up to ${\cal O} (10^{-6})$ at Belle II, over one order of magnitude improvement of $v_R$ up to ${\cal O}(10^7)$ GeV~\cite{DeLaCruz-Burelo:2020ozf} (see also Ref.~\cite{Guadagnoli:2021fcj} for weak prospect). 

Limits have also been obtained on the LFV tauon decays into ALP $a$ via $\tau \to \ell + a$ (with $\ell = e,\, \mu$) at proton beam dump experiments such as CHARM~\cite{Ema:2025bww}. After being produced, ALPs are assumed to decay into dilepton pairs $ee$, $e\mu$ and $\mu\mu$. The limits on ALP can be re-interpreted and recast onto the light $H_3$ (see Appendix~\ref{app:comparison} for comparison with ALP relevant processes). For simplicity it is assumed in Ref.~\cite{Ema:2025bww} that $g_{e\tau} = g_{\mu\tau}$ which is a good approximation in the LRSM. It turns out that the $v_R$ scale is excluded up to $5\times10^5$ GeV for scalar mass up to ${\cal O} (100)$ MeV. The precision of $\tau \to \ell + a$ can be improved by over one order of magnitude at future beam dump experiments such as SHiP~\cite{Ema:2025bww}, with $v_R$ pushed up to $2\times10^7$ GeV, as seen in Fig.~\ref{fig:LFV}.


The constraints from other LFV processes are rather weak, e.g. those from $\mu \to eee$ and $\mu \to e\gamma$, as these processes arise at the 2-loop or higher order in the LRSM if $H_3$ is involved. The limits from the meson decays $\pi^\pm,\, K^\pm \to \ell^\pm \nu H_3$, with $H_3$ emitted from charged leptons, are also rather weak, as the couplings can only constrained up to ${\cal O}(10^{-2})$~\cite{Guerrera:2022ykl,Dev:2024ygx,Jiang:2024cqj}.

\section{Astrophysical LFV constraints}

For a sub-GeV (pseudo)scalar, the most stringent astrophysical constraints on the LFV couplings are from the supernova observations.

{\it (i) SN1987A}.
In supernovae, ALP can be produced via the LFV coupling $g_{ae\mu}$ to electron and muon. When ALP is light, the production is dominated by muon decay $\mu \to e + a$. For a heavy ALP, the $e + \mu \to a$ process is more important. Within the mass range of roughly $(100, 110)$ MeV, the two processes above are kinematically forbidden or suppressed, and the lepton-proton bremsstrahlung process $\ell_\alpha + p \to \ell_\beta + p + a$ takes over to be the dominant one~\cite{Li:2025beu}. Benefiting from the high nucleon density in the supernova core, a sizable parameter space of ALP mass $m_a$ and $g_{ae\mu}$ is excluded by the observed neutrino luminosity of ${\cal L}_\nu \simeq 3\times 10^{52}$ erg/s of SN1987A~\cite{Li:2025beu} (see also Ref.~\cite{Zhang:2023vva}).  Recast onto the light $H_3$, the $v_R$ scale is constrained up to $2\times10^7$ GeV.

{\it (ii) low-energy supernovae}.  The low-energy supernovae (LESNe) have been identified, and the explosion energy can be as low as $10^{50}$ erg, which provides extra constraints on light particles~\cite{Chugai:1999en,Pastorello:2003tc,Burrows:2020qrp,Kitaura:2005bt,Melson:2015tia,Radice:2017ykv,Muller:2018utr,Burrows:2019rtd,Stockinger:2020hse}. Taking into account all the dominant production channels of $\mu \to e + a$, $e+\mu \to a$ and $\ell_\alpha + \gamma \to \ell_\beta +a$, the coupling $g_{ae\mu}$ is constrained up to ${\cal O}(10^{-11})$ for ALP mass up to roughly 550 MeV~\cite{Huang:2025rmy,Huang:2025xvo}. The corresponding constraint on the right-handed scale is $v_R \gtrsim 4\times10^8$ GeV, over one order of magnitude stronger than the SN1987A limit above.

There might also be some cosmological constraints on (pseudo)scalars below the GeV scale, e.g. those from the big bang nucleosynthesis and cosmic microwave background. However, these cosmological limits depend largely on the cosmological evolution history such as the reheating temperature~\cite{Cadamuro:2011fd,Berger:2016vxi,Hufnagel:2017dgo,Fradette:2018hhl,Depta:2020wmr,Depta:2020zbh,Balazs:2022tjl,Langhoff:2022bij,DEramo:2024lsk,Akita:2024nam,EscuderoAbenza:2025tsi,Jung:2025dyo}. Therefore, we will not consider the cosmological constraints in this work.

\section{LFC constraints}

The laboratory constraints on the LFC couplings are rather weak, e.g. those from the decays $\mu \to e \nu \bar\nu a$~\cite{Knapen:2024fvh}, $\tau \to \pi \nu a$~\cite{Ema:2025bww,Jiang:2025nie}, $\pi,\, K \to \ell \nu X$~\cite{Guerrera:2022ykl,Dev:2024ygx,Jiang:2024cqj}.
The most stringent constraint on the coupling $g_e$ of $H_3$ with electrons are from the supernova observations~\cite{Calibbi:2020jvd,Calore:2021klc,Ferreira:2022xlw,Carenza:2021pcm,Fiorillo:2025sln,Ferreira:2025qui}. The production of ALP in supernovae is dominated by the semi-Compton process $e+\gamma \to e + a$. When ALP is heavy, the process $e^+ + e^- \to a$ is also important. The decay of $a \to e^+ e^-$ in the mantle of LESNe is severely constrained by the limit of deposited energy of $10^{50}$ erg~\cite{Caputo:2022mah}, which excludes the value of $g_{ae}$ down to ${\cal O} (10^{-11})$~\cite{Fiorillo:2025sln}. Adopting conservatively the SFHo-18.8 supernova model~\cite{Page:2020gsx} and re-interpreting the limit (see Appendix~\ref{app:limit} for details), the $v_R$ scale is excluded up to $10^6$ GeV, as indicated by the dot-dashed magenta line in Fig.~\ref{fig:LFV}. There are also some other supernova limits on ALPs, e.g. those from $\gamma$-ray signals from SN1987A and the luminosity limits. However, the corresponding constraints on the $v_R$ scale are relatively weaker~\cite{Qiang:2026abc}.

The supernova constraints on the coupling of ALP to muons are also obtained~\cite{Croon:2020lrf,Calibbi:2020jvd,Ferreira:2025qui}. However, as the muon mass is significantly larger than the typical supernova temperature $T \sim 30$ MeV, the corresponding supernova limits on the couplings of $H_3$ to muons are expected to differ significantly from that on ALPs~\cite{Qiang:2026abc}. A dedicated analysis is necessary to obtain robust supernova limits on the coupling of $H_3$ to muons, which is far beyond the main scope of this paper.

\section{Parametric dependence of the LFV and LFC couplings and the $v_R$ constraints}

\begin{figure*}
    \centering
    \includegraphics[width= 0.45\linewidth]    {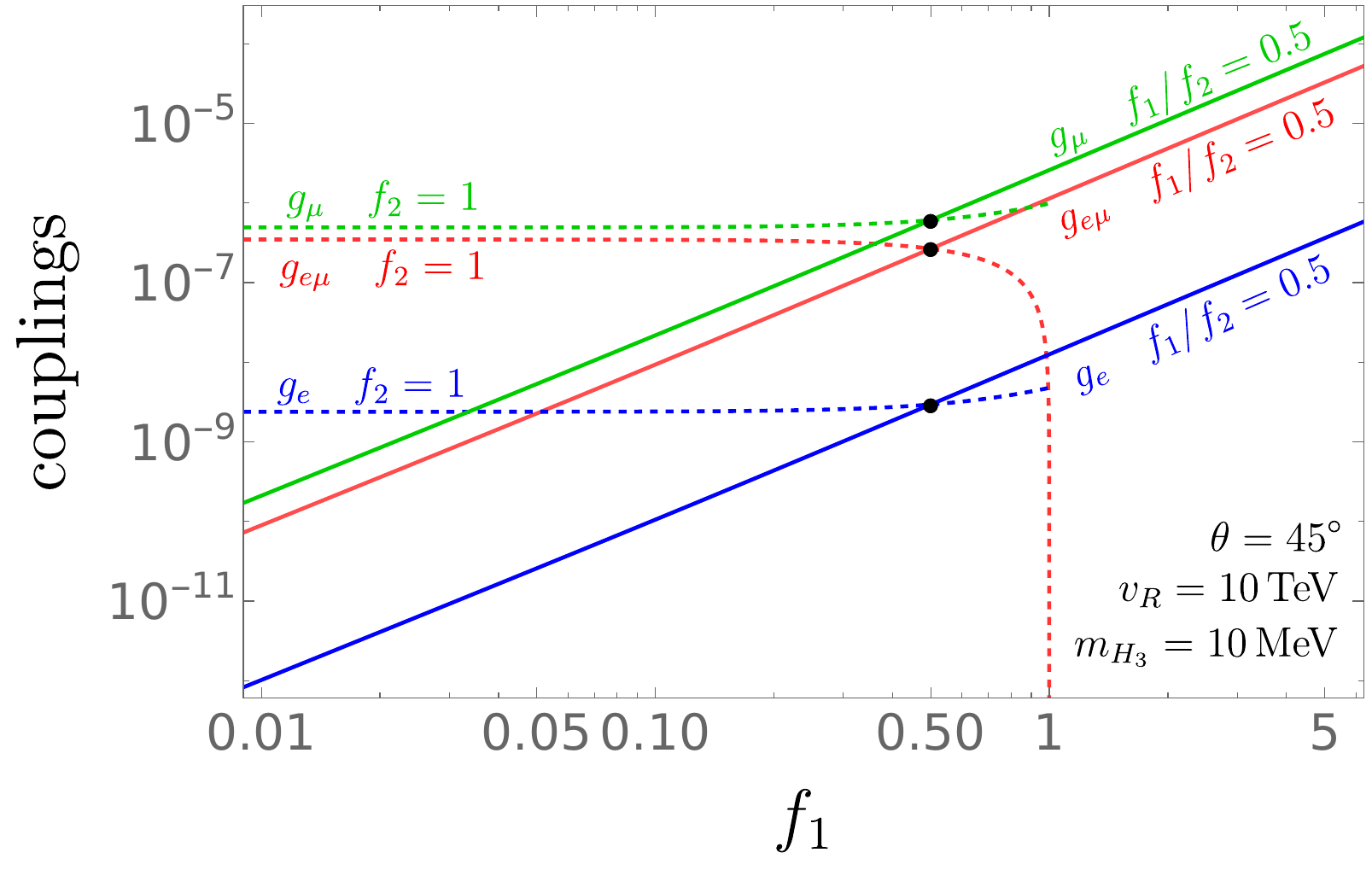}
    \includegraphics[width= 0.45\linewidth]    {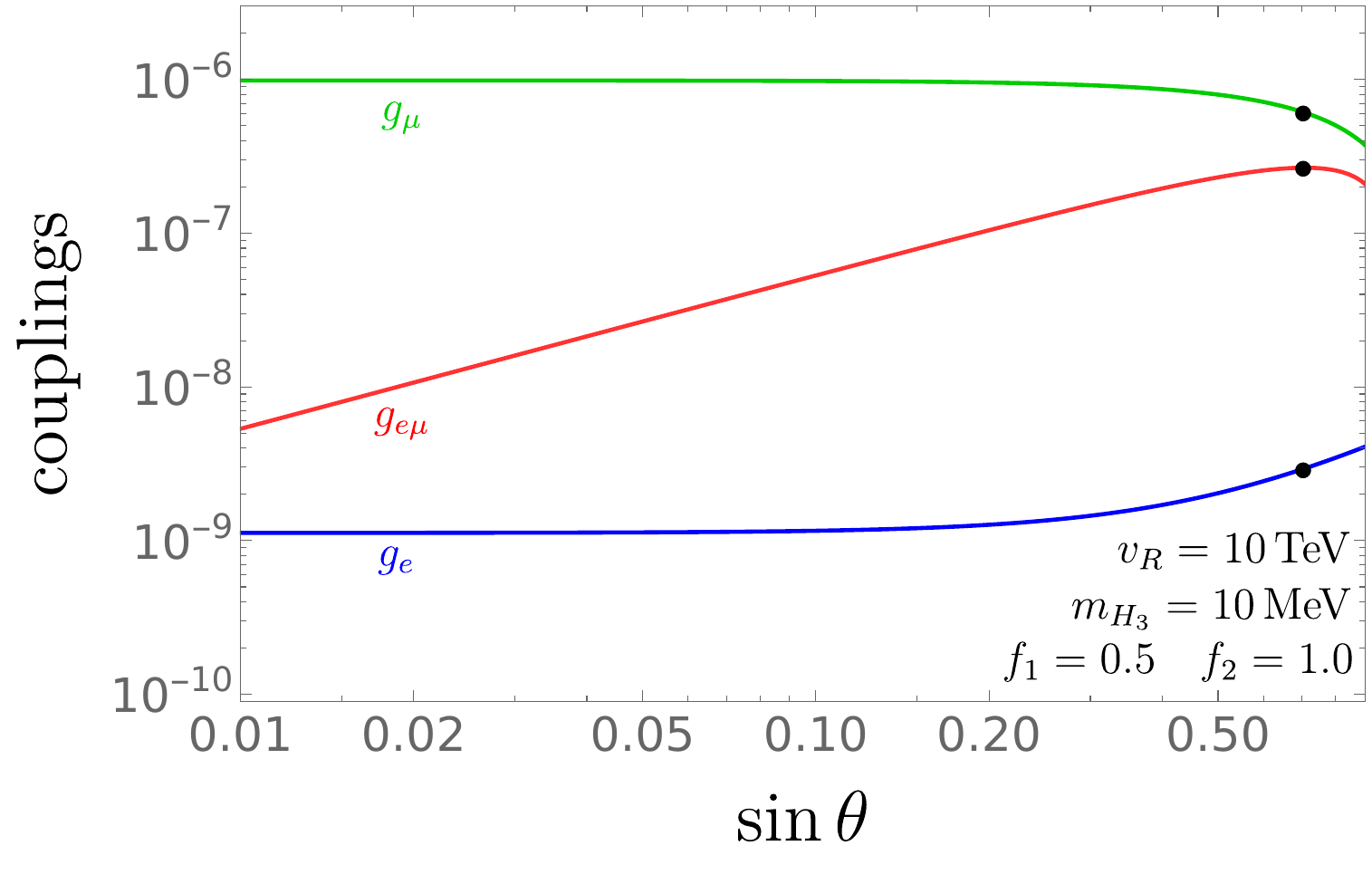}
    \caption{Parametric dependence of the effective LFV coupling $g_{e\mu}$ (red), and LFC couplings $g_e$ (blue) and $g_{\mu}$ (green) on the Yukawa coupling $f_1$ with $f_1/f_2 = 0.5$ (solid) or $f_2=1$ (dashed) in the left panel, and on the sine of RHN mixing angle $\sin\theta$ in the right panel. The benchmark point of $f_1=0.5$, $f_2=1.0$ is labeled by the black dots, and other parameters are fixed as shown in the plots.   }
    \label{fig:parameter:coupling}
\end{figure*}

We have taken the specific benchmark values for the Yukawa couplings $f_{1,\,2}$ and RHN mixing angle $\theta$. For other choices of these parameters, the LFV and LFC couplings and the resulting constraints on the $v_R$ scale could be very different. For illustration purpose, the parametric dependence of the effective  couplings $g_{e\mu}$, $g_{e}$ and $g_\mu$ on the Yukawa couplings $f_{1,\,2}$ and the RHN mixing angle is shown in Fig.~\ref{fig:parameter:coupling}. The benchmark point of $f_1 = 0.5$ and $f_2 = 1.0$ are labeled by the black dots in both the left and right panels.
\begin{itemize}
    \item {\it Dependence on $f_{1}$}. Let us first take the benchmark values of scalar mass $m_{H_3} = 10$ MeV, the scale $v_R = 10$ TeV and the mixing angle $\theta = 45^\circ$. Fixing the Yukawa coupling ratio $f_1/f_2 = 0.5$, and varying the parameter $f_1$, the resulting LFV coupling $g_{e\mu}$ and LFC couplings $g_{e}$, $g_{\mu}$ are shown as the solid red, blue and green lines in the left panel of Fig.~\ref{fig:parameter:coupling}. It turns out that these effective couplings are universally proportional to the Yukawa couplings via $\propto f_1 f_2$ in such a parameter setup.
    \item {\it Dependence on $f_{1}/f_2$}. Fixing $f_2 = 1.0$ and other parameters the same as above, let us vary only $f_1$ (keeping $f_1 \leq f_2$), which is equivalent to changing the ratio $f_1/f_2$. The corresponding couplings $g_{e\mu}$, $g_e$ and $g_\mu$ are presented as the dashed red, blue and green lines in the left panel of Fig.~\ref{fig:parameter:coupling}. As indicated by these lines, in the limit of $f_1 \ll f_2$, the LFV and LFC couplings are both dominated by the larger Yukawa coupling $f_2$ (or the largest coupling in the case of three coupling involved). In the limit of $f_1 = f_2$, the LFV coupling $g_{e\mu}$ vanishes, as in this case the two RHNs are indistinguishable from each other.
    \item {\it Dependence on $\sin\theta$}. Fixing again $m_{H_3} = 10$ MeV, $v_R = 10$ TeV, and the Yukawa couplings $f_1 = 0.5$, $f_2 = 1.0$, let us vary only the sine of the mixing angle $\sin\theta$. The resultant couplings $g_{e\mu}$, $g_e$ and $g_\mu$ are shown as the solid red, blue and green lines in the right panel of Fig.~\ref{fig:parameter:coupling}. As expected and explicitly shown in Eqs.~(\ref{eqn:cL1}) to (\ref{eqn:cR5}), the LFV coupling $g_{e\mu}$ depend on the mixing angle of RHNs via $\sin\theta \cos\theta$. The LFC couplings $g_{e}$ and $g_{\mu}$ are almost constants when the mixing angle $\theta$ is small, and get moderately altered when the mixing is significant, as implied by Eqs.~(\ref{eqn:cbeta:L1}) to (\ref{eqn:calpha:R2}).
\end{itemize}

\begin{figure*}
    \centering
    \includegraphics[width= 0.45\linewidth]{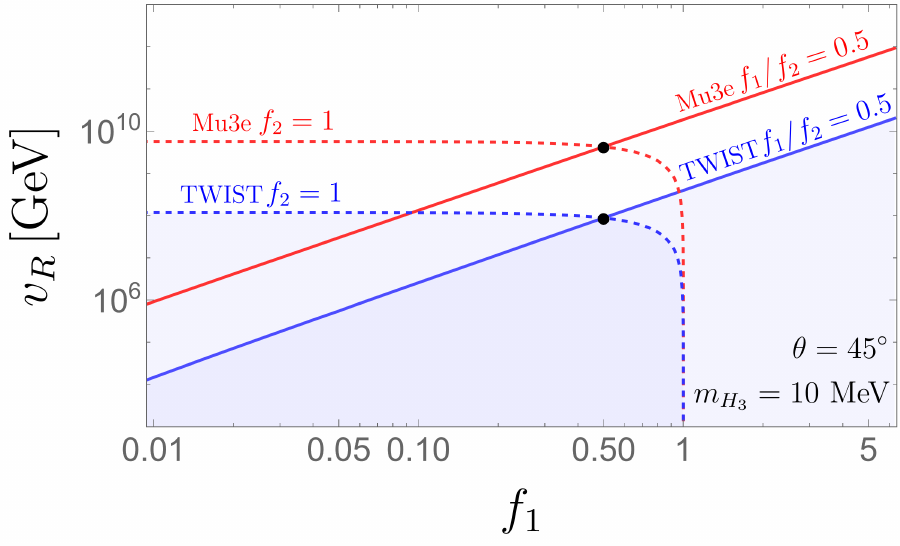}
    \includegraphics[width= 0.45\linewidth]{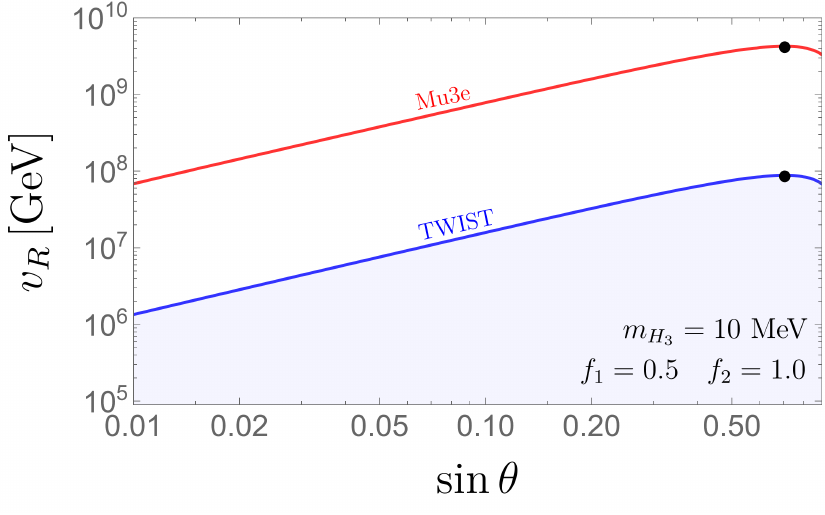}
    \caption{
    Parametric dependence of the TWIST limit on $v_R$ (shaded blue) and the future prospect of $v_R$ at the Mu3e experiment (red) on the Yukawa coupling $f_1$ with $f_1/f_2 = 0.5$ (solid) or $f_2=1$ (dashed) in the left panel, and on the sine of RHN mixing angle $\sin\theta$ in the right panel. The benchmark point of $f_1=0.5$, $f_2=1.0$ is labeled by the black dots, and other parameters are fixed as shown in the plots. }
    \label{fig:parameter:vR}
\end{figure*}

The resulting dependence of the $v_R$ constraints on the Yukawa couplings $f_{1,\,2}$ and the sine of RHN mixing angle $\sin\theta$ is presented in Fig.~\ref{fig:parameter:vR}. As the LFC limits are much weaker, for illustration purpose we choose the current LFV limit from TWIST and the future prospect of Mu3e in the channel of $\mu \to e + {\rm inv}$, which are presented in Fig.~\ref{fig:parameter:vR} as the shaded blue regions and the red lines, respectively. As in Fig.~\ref{fig:parameter:coupling}, the solid and dashed lines in the left panel are for the cases of fixed ratio $f_1/f_2 = 0.5$ and $f_2 = 1$, respectively, while the right panel is for the dependence on the sine of the RHN mixing angle $\sin\theta$. Other fixed parameters are shown in the plots, and the benchmark point of $f_1 = 0.5$ and $f_2 = 1.0$ are indicated by the black dots in both the two panels.
The corresponding partial width is proportional to the LFV coupling via $\Gamma (\mu \to e + H_3) \propto |g_{e\mu}|^2$, therefore the constraints on $v_R$ in Fig.~\ref{fig:parameter:vR} exhibit the same dependence on the Yukawa couplings $f_{1,\,2}$ and the sine of RHN mixing angle $\sin\theta$ as that for the LFV coupling $g_{e\mu}$ in Fig.~\ref{fig:parameter:coupling}. In particular, the constraints on $v_R$ will be enhanced when both $f_{1,\,2}$ are large (with the ratio $f_{1}/f_2$ fixed), and get weaker when both $f_{1,2}$ are small, $f_1 \to f_2$ or $\sin\theta$ is small. For instance, for the benchmark value of $f_1 = 0.5$, $f_2 = 1.0$ and other parameters the same as in the left panel of Fig.~\ref{fig:parameter:vR}, the TWIST limit (Mu3e prospect) on $v_R$ is $9.3\times10^7$ ($4.3\times10^9$) GeV; the limit will weaken to $2.7\times10^6$ ($1.4\times10^8$) GeV for the case of $f_1 = 0.1$, $f_2 = 0.2$, and to $1.9\times10^6$ ($9.5\times10^7$) GeV for the almost degenerate case of $f_1 = 0.99$, $f_2 = 1$. When the mixing angle $\sin\theta = 0.01$ but fixing other parameters as in the right panel of Fig.~\ref{fig:parameter:vR}, the TWIST limit (Mu3e prospect) on $v_R$ will decrease to $1.3\times10^6$ ($6.9\times10^7$) GeV.

\section{Diphoton constraints}

On the coupling of $H_3$ with diphoton, the most constraining laboratory limits
are from the beam dump experiments, e.g. the current limits from  E137~\cite{Liu:2023bby}, MiniBooNE~\cite{Capozzi:2023ffu}, CHARM~\cite{Dobrich:2019dxc}, NuCal~\cite{Dobrich:2019dxc} and the prospect of SHiP~\cite{Patrone:2025fwk,Dobrich:2019dxc} (see Refs.~\cite{Dobrich:2019dxc,Blinov:2021say,Patrone:2025fwk,NA64:2020qwq,Dusaev:2020gxi,Capozzi:2023ffu,Feng:2018pew,Bai:2021gbm,Alekhin:2015byh,Dobrich:2015jyk} for weaker prospects). However, it turns out these constraints are not stringent enough to set limits on $v_R$ above 10 TeV~\cite{Qiang:2026abc}.

In comparison, the astrophysical constraints on the coupling of ALP to photons are much more stringent (collected in Fig.~\ref{fig:LFV} and Table~\ref{tab:gauge}).
The production of ALPs inside supernovae via the coupling $g_{a\gamma\gamma}$ is dominated by the Primakoff process $\gamma + {\cal N} \to {\cal N} + a$ and photon coalescence $\gamma + \gamma \to a$. The $\gamma$-rays generated from ALP decay from SN1987A, SN2023ixf and type Ic supernovae exclude the coupling $g_{a\gamma\gamma}$ up to ${\cal O} (10^{-12})$ ${\rm GeV}^{-1}$~\cite{Jaeckel:2017tud,Hoof:2022xbe,Muller:2023vjm,Muller:2023pip,Diamond:2023scc,Candon:2025ypl}. The energy deposition of ALPs in LESNe~\cite{Caputo:2022mah,Fiorillo:2025yzf} and the $X$-ray and $\gamma$-ray observations of the neutron star (NS) merger GW170817~\cite{Diamond:2023cto,Dev:2023hax} provide extra limits, with the coupling $g_{a\gamma\gamma}$ constrained up to ${\cal O}(10^{-10})$ GeV$^{-1}$. 
Depending on its mass, the light scalar $H_3$ can decay into $\gamma\gamma$ or dileptons $e^+ e^-$, $e^\pm \mu^\mp$, $\mu^+ \mu^-$. The leptons from $H_3$ decay are very likely to annihilate into photons via the process $\ell^+ + \ell^- \to \gamma + \gamma$. For simplicity, we assume all the decay products of $H_3$ convert finally into $\gamma$-rays. As seen in Fig.~\ref{fig:LFV}, the $v_R$ scale is excluded up to $2\times10^9$ GeV.
The sensitivity of $g_{a\gamma\gamma}$ can be largely improved up to $10^{-14}$, if supernova explosion happens at the red supergiant Betelgeuse, which is only $\sim 200$ pc away~\cite{Jaeckel:2017tud}.
The future observation of $\gamma$-rays from NS mergers in the next generations telescopes can also explore more open parameter space of $g_{a\gamma\gamma}$~\cite{Dev:2023hax}. The corresponding prospects of $v_R$ can reach up to $6\times10^{11}$ GeV, which is expected to be the highest probable range of $v_R$, as shown in Fig.~\ref{fig:LFV}.

\section{Discussions and conclusion}

All the current leading LFV, LFC and diphoton constraints on $m_{H_3}$ and $v_R$ are presented as the shaded regions in Fig.~\ref{fig:LFV}, and the predominant future prospects in various laboratory experiments and astrophysical observations are depicted as lines. The LFV, LFC and diphoton constraints are labeled as the solid, dot-dashed and dashed lines, respectively. The $\gamma$-rays signals from SN1987A induced by the diphoton coupling provide the current most stringent limit on $v_R$, up to $2\times10^9$ GeV, which can be improved up to $6\times10^{11}$ GeV by future observations of Betelgeuse.
The baryon asymmetry of the Universe can be naturally explained by the decay of heavy neutrinos via the leptogenesis mechanism~\cite{Fukugita:1986hr}. In the framework of LRSM, it is required that $v_R \gtrsim {\cal O}(10^{11})$ GeV if the heavy neutrinos have the hierarchical mass spectrum~\cite{Frere:2008ct,ReFiorentin:2016rzn}. It is remarkable that the future astrophysical observations could reach the non-resonant leptogenesis scale in the LRSM. In absence of fortune to see a nearby supernova explosion in the near future, the searches of the LFV signal $\mu \to e + {\rm inv}$ in the Mu3e experiment can improve the current SN1987A limit by roughly a factor of $2.5$, up to roughly $5\times10^9$ GeV, if the Yukawa couplings $f_i$ are of order one and the RHN mixing is maximal.

A relatively heavier $H_3$ can be directly produced at future high-energy lepton colliders via its coupling to photons and leptons, e.g. $e^+ + e^- \to \gamma + H_3$~\cite{Bauer:2018uxu,Mimasu:2014nea,Jaeckel:2015jla,Wang:2022ock,Schafer:2022shi,Bao:2025tqs}. Depending on the mass $m_{H_3}$, the signals may be prompt-like or displaced vertices. In analogy to the coupling $H_3 \gamma\gamma$, $H_3$ can also have the coupling to a photon and a $Z$ boson, i.e. in the form of $H_3 \gamma Z$, which is induced by the heavy charged scalars and $W_R$ boson~\cite{Nemevsek:2016enw}. Then we can expect some rare $Z$ decays such as $Z \to \gamma + H_3 \to 3\gamma$ at future Tera-$Z$ factories~\cite{Jaeckel:2015jla,Bauer:2017ris,Brivio:2017ije,Bauer:2018uxu}. In addition, $H_3$ can also have flavor-changing neutral current (FCNC) couplings to quarks, which are induced by the right-handed gauge couplings of $W_R$ boson. However, the FCNC decays $d_j \to d_i + H_3$ (with $d_{i,j}$ the down-type quarks) are heavily suppressed by the right-handed scale via $v_R^{-6}$ and the corresponding limits are rather weak~\cite{Qiang:2026abc}.

All the astrophysical constraints on the couplings of $H_3$ to charged leptons and photons are from the corresponding ALP limits in the literature, and the conversion of the LFV, LFC and diphoton couplings are performed separately. In light of the well-defined ultraviolet structure of the minimal LRSM, a combined analysis of all these channels in the framework of LRSM is phenomenologically in demand for more robust astrophysical constraints on the LRSM. Furthermore, it is very likely that such dedicated analysis could explore even higher values of the right-handed scale $v_R$.

In summary, we have performed a detailed study on the 1-loop couplings of light $SU(2)_R$-breaking scalar $H_3$ in the minimal LRSM, and focused mainly on the radiative LFV, LFC and diphoton couplings of $H_3$. In the limit of decoupling left-handed triplet $\Delta_L$, the LFV couplings originate from the mixing of heavy RHNs.
These LFV, LFC and diphoton couplings are severely constrained by the high-precision laboratory experiments and astrophysical observations. The current most stringent limit is from the $\gamma$-ray observations of SN1987A, which exclude the right-handed scale $v_R$ up to $2\times10^9$ GeV. Improved precision and unprecedented amount of data in future high-precision muon experiments and supernova observations can push the right-handed scale up to $5\times10^9$ GeV (assuming the RHNs are at the order of the $v_R$ scale and the RHN mixing is maximal) and $6\times10^{11}$ GeV, respectively.

\section*{Acknowledgments}


The authors are also grateful to Rabindra Mohapatra, P. S. Bhupal Dev, Sudip Jana, Zuowei Liu and Zeren S. Wang for the enlightening discussions, and Michael J. Ramsey-Musolf, Xiao-Gang He, Shao-Feng Ge and Shu Li for the valuable comments. S.Q. and Y.Z. are supported by the National Natural Science Foundation of China under grant No. 12175039,
the State Key Laboratory of Dark Matter Physics,
and the ``Fundamental Research Funds for the Central Universities''.
P.W. acknowledges support from Natural Science Foundation of Jiangsu Province (Grant No. BK20210201), Fundamental Research Funds for the Central Universities, Excellent Scholar Project of Southeast University (Class A), and the Big Data Computing Center of Southeast University.


\section*{Data availability}

The data for Fig.~\ref{fig:LFV} of this paper are openly available at Ref.~\cite{data}. More data are available from the authors upon reasonable request.


\appendix


\section{Partial widths for the LFV, LFC and diphoton decays}
\label{app:width}

The expressions of the LFV couplings $c_{\alpha\beta}^{(L,R)}$ in Eq.~(\ref{eqn:LFV}) can be found in Eq.~(\ref{eqn:cLcR}). With these couplings,
it is trivial to get the partial width for the LFV decay of $H_3$:
\begin{widetext}
\begin{eqnarray}
\label{eqn:width:LFV}
\Gamma (H_3 \to \ell_\alpha^\pm \ell_\beta^\mp) &\equiv&
\Gamma (H_3 \to \ell_\alpha^+ \ell_\beta^-) + \Gamma (H_3 \to \ell_\alpha^- \ell_\beta^+) \nonumber \\
&=& \frac{m_{H_3}}{8\pi} \left[ \left( \left| c_{\alpha\beta}^{(L)} \right|^2 + \left| c_{\alpha\beta}^{(R)} \right|^2 \right) (1-\eta_\alpha - \eta_\beta) - 4 {\rm Re} \left( c_{\alpha\beta}^{(L)} c_{\alpha\beta}^{(R)\ast} \right) \sqrt{\eta_\alpha \eta_\beta} \right] \nonumber \\
&& \times \Big[ \Big( 1- (\sqrt{\eta_\alpha} + \sqrt{\eta_\beta})^2 \Big) \Big( 1- (\sqrt{\eta_\alpha} - \sqrt{\eta_\beta})^2 \Big) \Big]^{1/2}  \,,
\end{eqnarray}
with $\eta_\alpha \equiv m_{\ell_\alpha}^2/m_{H_3}^2$.
For the case of $m_{H_3} < m_{\ell_\beta} - m_{\ell_\alpha}$, the decay $\ell_\beta \to \ell_\alpha H_3$ is kinematically allowed, and the corresponding partial width is
\begin{eqnarray}
\label{eqn:width:LFV2}
\Gamma (\ell_\beta \to \ell_\alpha H_3) &=& \frac{m_{\ell_\beta}}{32\pi} \left[ \left( \left|c_{\alpha\beta}^{(L)}\right|^2 + \left|c_{\alpha\beta}^{(R)}\right|^2 \right) (1-\lambda_\alpha - \lambda_{H_3}) + 4 {\rm Re} \left(c_{\alpha\beta}^{(L)} c_{\alpha\beta}^{(R)\ast}\right) \sqrt{\lambda_\alpha} \right] \nonumber \\
&& \times \Big[ \Big( 1- (\sqrt{\lambda_\alpha} + \sqrt{\lambda_{H_3}})^2 \Big) \Big( 1- (\sqrt{\lambda_\alpha} - \sqrt{\lambda_{H_3}})^2 \Big) \Big]^{1/2}  \,,
\end{eqnarray}
\end{widetext}
where $\lambda_X \equiv m_X^2 / m_{\ell_\beta}^2$. For the LFC couplings, the partial width reads
\begin{equation}
\Gamma (H_3 \to \ell^+ \ell^-) = \frac{g_\ell^2 m_{H_3}}{8\pi} \left( 1 - 4\eta_\ell \right)^{3/2} \,,
\end{equation}
where the expression for the coupling $g_\ell$ is given in Eq.~(\ref{eqn:gl}).

The coupling of $H_3$ with diphoton are induced at the 1-loop order by the singly-charged scalar $H_1^\pm$ from the bidoublet $\Phi$, the singly-charged scalar $H_L^\pm$ and doubly-charged scalar $H_L^{\pm\pm}$ from the left-handed triplet $\Delta_L$, and the doubly-charged scalar $H_R^{\pm\pm}$ from the right-handed triplet $\Delta_R$. In the limit of vanishing mixing of $H_3$ with the SM Higgs and other heavy scalars in the LRSM, the partial width for the diphoton decay channel is~\cite{BhupalDev:2016nfr,Dev:2017dui,Nemevsek:2016enw}
\begin{eqnarray}
\label{eqn:diphoton}
\Gamma (H_3 \to \gamma \gamma) &=& \frac{m_{H_3}^3}{32 \pi v_R^2} \frac{\alpha^2}{(4 \pi)^2}  \left| F \right|^2 \,.
\end{eqnarray}
The loop function is
\begin{eqnarray}
\label{eqn:F}
F &=& A_0 (\eta_{H_1^\pm}) + A_0 (\eta_{H_L^\pm}) + 4 A_0 (\eta_{H_L^{\pm\pm}}) \nonumber \\
&&  + 4 A_0 (\eta_{H_R^{\pm\pm}}) + A_1 (\eta_{W_R^{}}) \,,
\end{eqnarray}
where the factor of $4$ is from the square of charges for the doubly-charged scalars, and the loop functions
\begin{eqnarray}
A_{0} (x) &=& - x \left[1 - x f(x) \right] \,, \\
A_1 (x) &=& - \left[2 + 3 x \left(1 + (2 - x) f(x) \right) \right] \,,
\end{eqnarray}
with
\begin{align}
\label{eqn:ftau}
f(x) \ \equiv \ \left\{ \begin{array}{ll}
{\rm \arcsin}^2\sqrt{x^{-1}}, & \tau \geq 1 \,, \\
-{\displaystyle \frac{1}{4}}\left[\log \left( \frac{1+\sqrt{1-x}}{1-\sqrt{1-x}}\right)-i\pi  \right]^2, & \tau<1 \,.
\end{array}\right.
\end{align}
In the limit of $m_{H_3} \to 0$, the loop functions
\begin{equation}
A_0 (\infty) = \frac13 \,, \qquad
A_1 (\infty) = -7 \,,
\end{equation}
which lead to
\begin{equation}
F \cong (1+1+4+4)\times \frac13 + (-7) = -\frac{11}{3} \,.
\end{equation}

\begin{figure*}
    \centering
    \includegraphics[width= 0.4\linewidth]{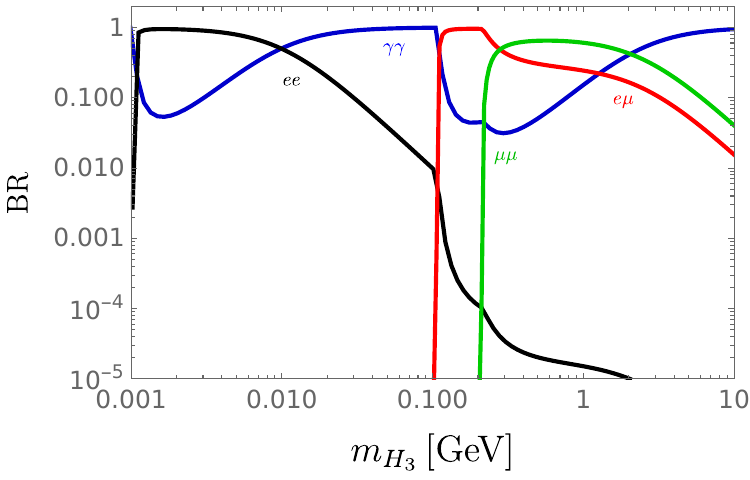}
    \includegraphics[width= 0.4\linewidth]{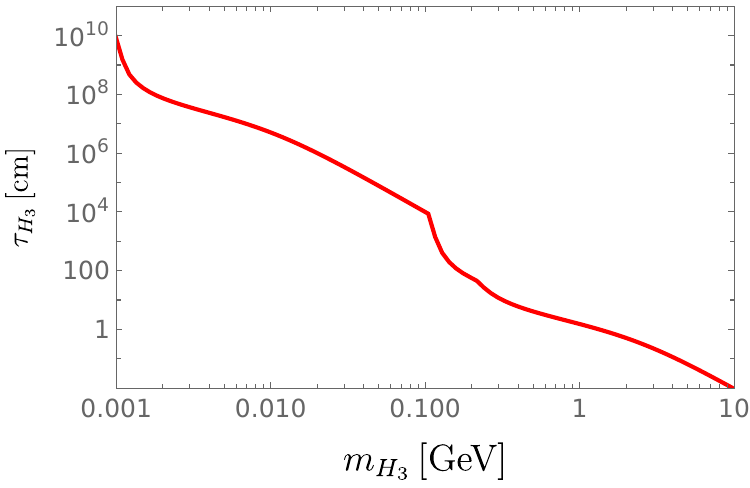}
    \caption{{\it Left panel}: The BRs of $H_3$ decaying into $e^+e^-$, $e^\pm\mu^\mp$, $\mu^+\mu^-$ and $\gamma\gamma$ as functions of its mass $m_{H_3}$. {\it Right panel}: The proper lifetime of $H_3$ as function of its mass $m_{H_3}$. We have set $v_R = 10$ TeV. Other parameter setups can be found in the main text.
    }
    \label{fig:BR}
\end{figure*}

For illustration purpose, the decay BRs of $H_3 \to e^+ e^-,\, e^\pm \mu^\mp,\, \mu^+ \mu^-,\, \gamma\gamma$ as functions of $m_{H_3}$ are shown in the left panel of Fig.~\ref{fig:BR}. The proper lifetime of $H_3$ is presented in the right panel of Fig.~\ref{fig:BR}, with the benchmark value of $v_R = 10$ TeV.

\section{Functions for the 1-loop LFV and LFC couplings}
\label{app:functions}

\subsection{LFV couplings}

In this subsection, we collect all the loop functions for the LFV couplings in Eq.~(\ref{eqn:LFV}), with the same notation convention of the A, B, C loop functions as in Ref.~\cite{Hahn:1998yk}. The effective couplings in Eq.~(\ref{eqn:LFV}) can be decomposed into the following contributions:
\begin{equation}
\label{eqn:cLcR}
c_{\alpha\beta}^{(X)} = \sum_{i=1}^{6} c_{\alpha\beta}^{(X),i} \,,
\end{equation}
with $X = L,\,R$. In the first diagram in Fig.~\ref{fig:diagram} with a right-handed doubly-charged scalar $H_R^{\pm\pm}$, the charged lepton flavor $\gamma = \alpha$ or $\gamma = \beta$ (assuming $\alpha \neq \beta$) in the case of mixing of two RHNs. For the cases of $\gamma = \alpha$ and $\gamma = \beta$, the effective couplings are labeled with the superscripts 1 and 2, respectively:
\begin{widetext}
\begin{eqnarray}
\label{eqn:cL1}
&& \left(\frac{m_\beta\sin\theta \cos\theta }{32 \sqrt{2} \pi ^2   v_R ^3}\right)^{-1}{c_{\alpha\beta}^{(L),1}} \ {\text{=}} \ \nonumber \\
&& -8   m_{H_R^{\pm\pm}} ^2 (  m_{N_1} -  m_{N_2} ) ((  m_{N_1} -  m_{N_2} ) \cos{2\theta} +  m_{N_1} +  m_{N_2} )  {\rm{C_1}}  (   m_{H_3} ^2,  m_\beta ^2,  m_\alpha ^2,  m_{H_R^{\pm\pm}} ^2,  m_{H_R^{\pm\pm}} ^2,  m_\alpha ^2  )  \,, \\
&& \left(\frac{m_\alpha\sin\theta \cos\theta}{32 \sqrt{2} \pi ^2   v_R ^3}\right)^{-1}{c_{\alpha\beta}^{(R),1}}\ {\text{=}} \ \nonumber \\
&&     16   m_{H_R^{\pm\pm}} ^2 (  m_{N_1} -  m_{N_2} ) \sin ^2  \theta   (  m_{N_1}  \cot ^2  \theta +  m_{N_2}   )   \left[ {\rm{C_0}}  (  m_{H_3} ^2,  m_\beta ^2,  m_\alpha ^2,  m_{H_R^{\pm\pm}} ^2,  m_{H_R^{\pm\pm}} ^2,  m_\alpha ^2  ) \right. \nonumber \\
&& \left. + {\rm{C_1}}  (  m_{H_3} ^2,  m_\beta ^2,  m_\alpha ^2,  m_{H_R^{\pm\pm}} ^2,  m_{H_R^{\pm\pm}} ^2,  m_\alpha ^2  )+ {\rm{C_2}}  (   m_{H_3} ^2,  m_\beta ^2,  m_\alpha ^2,  m_{H_R^{\pm\pm}} ^2,  m_{H_R^{\pm\pm}} ^2,  m_\alpha ^2  )  \right]   \,, \\
&& \left(\frac{m_\beta\sin\theta \cos\theta}{32 \sqrt{2} \pi ^2   v_R ^3}\right)^{-1} c_{\alpha\beta}^{(L),2} \ {\text{=}} \ \nonumber \\
&&  -16   m_{H_R^{\pm\pm}} ^2 (  m_{N_1} -  m_{N_2} ) \sin ^2  \theta   (  m_{N_1} +  m_{N_2}  \cot ^2  \theta  )  {\rm{C_1}}  (   m_{H_3} ^2,  m_\beta ^2,  m_\alpha ^2,  m_{H_R^{\pm\pm}} ^2,  m_{H_R^{\pm\pm}} ^2,  m_\beta ^2  )   \,, \\
&& \left(\frac{m_\alpha\sin\theta \cos\theta}{32 \sqrt{2} \pi ^2   v_R ^3}\right)^{-1} c_{\alpha\beta}^{(R),2} \ {\text{=}} \ \nonumber \\
&&  16   m_{H_R^{\pm\pm}} ^2 (  m_{N_1} -  m_{N_2} ) \sin ^2\theta   (  m_{N_1} +  m_{N_2}  \cot ^2\theta  )   \left[ {\rm{C_0}}  (   m_{H_3} ^2,  m_\beta ^2,  m_\alpha ^2,  m_{H_R^{\pm\pm}} ^2,  m_{H_R^{\pm\pm}} ^2,  m_\beta ^2  ) \right. \nonumber \\
&& \left. + {\rm{C_1}}  (  m_{H_3} ^2,  m_\beta ^2,  m_\alpha ^2,  m_{H_R^{\pm\pm}} ^2,  m_{H_R^{\pm\pm}} ^2,  m_\beta ^2  )+ {\rm{C_2}}  (  m_{H_3} ^2,  m_\beta ^2,  m_\alpha ^2,  m_{H_R^{\pm\pm}} ^2,  m_{H_R^{\pm\pm}} ^2,  m_\beta ^2  )  \right]    \,,
\end{eqnarray}
where $m_{\alpha,\beta}$ are the masses of the charged leptons with flavors $\alpha$ and $\beta$, respectively.

For the second diagram in Fig.~\ref{fig:diagram}, the couplings for the case with the RHN $N_1$ running in the loop are:
\begin{eqnarray}
&& \left(\frac{m_\beta\sin\theta \cos\theta}{32 \sqrt{2} \pi ^2   v_R ^3}\right)^{-1} c_{\alpha\beta}^{(L),3} \ \simeq \ \nonumber \\
&&    -6   m_{N_1} ^2  {\rm{B_{0}}}  (   m_\beta ^2,  m_{N_1} ^2,  m_{W_R} ^2  )-4   m_{N_1} ^2  {\rm{B_{1}}}  (   m_\beta ^2,  m_{N_1} ^2,  m_{W_R} ^2  )-2   m_{N_1} ^2   ( 
m_{N_1} ^2-2   m_{W_R} ^2  )  {\rm{C_0}}  (   m_{H_3} ^2,  m_\beta ^2,  m_\alpha ^2,  m_{N_1} ^2,  m_{N_1} ^2,  m_{W_R} ^2  )\nonumber \\
&&+4   m_{N_1} ^2  {\rm{C_{00}}}  (   m_{H_3} ^2,  m_\beta ^2,  m_\alpha ^2,  m_{N_1} ^2,  m_{N_1} ^2,  m_{W_R} ^2  )
+4   m_{N_1} ^2
(  
m_{N_1} ^2 +2   m_{W_R} ^2  )  {\rm{C_1}}  (   m_{H_3} ^2,  m_\beta ^2,  m_\alpha ^2,  m_{N_1} ^2,  m_{N_1} ^2,  m_{W_R} ^2  )\nonumber \\
&&+4   m_{H_3} ^2   m_{N_1} ^2  {\rm{C_{11}}}  (   m_{H_3} ^2,  m_\beta ^2,  m_\alpha ^2,  m_{N_1} ^2,  m_{N_1} ^2,  m_{W_R} ^2  )
-2   m_{N_1} ^2   (-  m_{H_3} ^2-  m_\alpha ^2+  m_\beta ^2  )  {\rm{C_{12}}}  (   m_{H_3} ^2,  m_\beta ^2,  m_\alpha ^2,  m_{N_1} ^2,  m_{N_1} ^2,  m_{W_R} ^2  )\nonumber \\
&&-4   m_\alpha ^2   m_{N_1} ^2  {\rm{C_{2}}}  (   m_{H_3} ^2,  m_\beta ^2,  m_\alpha ^2,  m_{N_1} ^2,  m_{N_1} ^2,  m_{W_R} ^2  )    \,, \\
&& \left(\frac{m_\alpha\sin\theta \cos\theta}{32 \sqrt{2} \pi ^2   v_R ^3}\right)^{-1} c_{\alpha\beta}^{(R),3} \ \simeq \ \nonumber \\
&&    -2   m_{N_1} ^2  {\rm{B_{0}}}  (   m_\beta ^2,  m_{N_1} ^2,  m_{W_R} ^2  )-2   m_{N_1} ^2   ( 
3   m_{N_1} ^2+2   m_{W_R} ^2  )  {\rm{C_0}}  (   m_{H_3} ^2,  m_\beta ^2,  m_\alpha ^2,  m_{N_1} ^2,  m_{N_1} ^2,  m_{W_R} ^2  )\nonumber \\
&&-4   m_{N_1} ^2  {\rm{C_{00}}}  (   m_{H_3} ^2,  m_\beta ^2,  m_\alpha ^2,  m_{N_1} ^2,  m_{N_1} ^2,  m_{W_R} ^2  )-2   m_{N_1} ^2   (  m_{H_3} ^2+  m_\alpha ^2-  m_\beta ^2  )  {\rm{C_{22}}}  (   m_{H_3} ^2,  m_\beta ^2,  m_\alpha ^2,  m_{N_1} ^2,  m_{N_1} ^2,  m_{W_R} ^2  ) \nonumber \\
&&-4   m_{H_3} ^2   m_{N_1} ^2  {\rm{C_{11}}}  (   m_{H_3} ^2,  m_\beta ^2,  m_\alpha ^2,  m_{N_1} ^2,  m_{N_1} ^2,  m_{W_R} ^2  )-2   m_{N_1} ^2   (3   m_{H_3} ^2+  m_\alpha ^2-  m_\beta ^2  )  {\rm{C_{12}}}  (   m_{H_3} ^2,  m_\beta ^2,  m_\alpha ^2,  m_{N_1} ^2,  m_{N_1} ^2,  m_{W_R} ^2  )\nonumber \\
&& -4   m_{N_1} ^2   (
 m_{N_1} ^2+2   m_{W_R} ^2  ) \left[ {\rm{C_1}}  (   m_{H_3} ^2,  m_\beta ^2,  m_\alpha ^2,  m_{N_1} ^2,  m_{N_1} ^2,  m_{W_R} ^2  )+ {\rm{C_{2}}}  (   m_{H_3} ^2,  m_\beta ^2,  m_\alpha ^2,  m_{N_1} ^2,  m_{N_1} ^2,  m_{W_R} ^2  ) \right] 
\,.
\end{eqnarray}
Here we have neglected the terms of $m_{H_3,\, \alpha,\, \beta} \ll m_{W_R,\, N_1,\, N_2}$ for example $m_{\alpha}^2 + m_{W_R}^2 \simeq m_{W_R}^2$. The corresponding couplings $c_{\alpha\beta}^{(L),4}$ and $c_{\alpha\beta}^{(R),4}$ for the diagram with the RHN $N_2$ can be easily obtained by substituting \(m_{N_1}\) with \(m_{N_2}\), and replacing \(\sin\theta\) with \(-\sin\theta\) in the prefactor.

For the last diagram in Fig.~\ref{fig:diagram}, if the RHN $N_1$ is in the propagator, the couplings are:
\begin{eqnarray}
&& \left(\frac{m_\beta\sin\theta \cos\theta}{32 \sqrt{2} \pi ^2   v_R ^3}\right)^{-1} c_{\alpha\beta}^{(L),5} \ \simeq  \nonumber \\
&&   2  (
   m_{N_1} ^2-  m_{W_R} ^2  )  {\rm{B_{0}}}  (   m_\beta ^2,  m_{N_1} ^2,  m_{W_R} ^2  ) +2  {\rm{B_{00}}}  (   m_\beta ^2,  m_{N_1} ^2,  m_{W_R} ^2  )+2   (
m_{N_1} ^2-  m_{W_R} ^2  )  {\rm{B_{1}}}  (   m_\beta ^2,  m_{N_1} ^2,  m_{W_R} ^2  ) \nonumber \\
&&+  (  m_{H_3} ^2-  m_\alpha ^2+  m_\beta ^2  )  {\rm{B_{11}}}  (   m_\beta ^2,  m_{N_1} ^2,  m_{W_R} ^2  ) +  m_{W_R} ^2(
3   m_\alpha ^2  +  m_\beta ^2   -  m_{H_3} ^2     )  {\rm{C_{12}}}  (   m_\alpha ^2,  m_{H_3} ^2,  m_\beta ^2,  m_{N_1} ^2,  m_{W_R} ^2,  m_{W_R} ^2  )\nonumber \\
&& 
-2   m_{W_R} ^4  {\rm{C_{0}}}  (   m_\alpha ^2,  m_{H_3} ^2,  m_\beta ^2,  m_{N_1} ^2,  m_{W_R} ^2,  m_{W_R} ^2  )
+  
2   m_{W_R} ^2   {\rm{C_{00}}}  (   m_\alpha ^2,  m_{H_3} ^2,  m_\beta ^2,  m_{N_1} ^2,  m_{W_R} ^2,  m_{W_R} ^2  )\nonumber \\
 &&-4   m_\alpha ^2  {\rm{C_{001}}}  (   m_\alpha ^2,  m_{H_3} ^2,  m_\beta ^2,  m_{N_1} ^2,  m_{W_R} ^2,  m_{W_R} ^2  ) +  4(   m_{H_3} ^2-  m_\beta ^2  )  {\rm{C_{002}}}  (   m_\alpha ^2,  m_{H_3} ^2,  m_\beta ^2,  m_{N_1} ^2,  m_{W_R} ^2,  m_{W_R} ^2  )\nonumber \\
&& +  m_{W_R} ^2 (
  m_{H_3} ^2  +3   m_\alpha ^2   -  m_\beta ^2    )  {\rm{C_{1}}}  (   m_\alpha ^2,  m_{H_3} ^2,  m_\beta ^2,  m_{N_1} ^2,  m_{W_R} ^2,  m_{W_R} ^2  )\nonumber \\
&& +  m_\alpha ^2 (  m_{H_3} ^2  +  m_\alpha ^2-     m_\beta ^2  ) \left[ {\rm{C_{11}}}  (   m_\alpha ^2,  m_{H_3} ^2,  m_\beta ^2,  m_{N_1} ^2,  m_{W_R} ^2,  m_{W_R} ^2  ) +2
{\rm{C_{112}}}  (   m_\alpha ^2,  m_{H_3} ^2,  m_\beta ^2,  m_{N_1} ^2,  m_{W_R} ^2,  m_{W_R} ^2  ) \right] \nonumber \\
&& -( (m_{H_3} ^2-m_\beta ^2)^2 + m_\alpha^2 (2 m_{H_3}^2 - 3 m_\alpha^2 + 2 m_\beta ^2))
{\rm{C_{122}}}  (   m_\alpha ^2,  m_{H_3} ^2,  m_\beta ^2,  m_{N_1} ^2,  m_{W_R} ^2,  m_{W_R} ^2  )\nonumber \\
&& 
-6   m_{W_R} ^4  {\rm{C_{2}}}  (   m_\alpha ^2,  m_{H_3} ^2,  m_\beta ^2,  m_{N_1} ^2,  m_{W_R} ^2,  m_{W_R} ^2  ) +  (  (  m_{H_3} ^2-  m_\alpha ^2  )^2-  m_\beta ^4  )  {\rm{C_{222}}}  (   m_\alpha ^2,  m_{H_3} ^2,  m_\beta ^2,  m_{N_1} ^2,  m_{W_R} ^2,  m_{W_R} ^2  )\nonumber \\
&& +  m_{W_R} ^2 (
  m_\alpha ^2    +3   m_\beta ^2-  m_{H_3} ^2        )  {\rm{C_{22}}}  (   m_\alpha ^2,  m_{H_3} ^2,  m_\beta ^2,  m_{N_1} ^2,  m_{W_R} ^2,  m_{W_R} ^2  )
\,, \\
\label{eqn:cR5}
&& \left(\frac{m_\alpha\sin\theta \cos\theta}{32 \sqrt{2} \pi ^2   v_R ^3}\right)^{-1} c_{\alpha\beta}^{(R),5} \ \simeq \ \nonumber \\
 &&  2  {A_0}  (  m_{W_R} ^2 ) + 2 (
    m_{N_1} ^2-   m_{W_R} ^2  )  {\rm{B_{0}}}  (   m_\beta ^2,  m_{N_1} ^2,  m_{W_R} ^2  ) \nonumber \\
 && -2  {\rm{B_{00}}}  (   m_\beta ^2,  m_{N_1} ^2,  m_{W_R} ^2  )+  (  m_{H_3} ^2-  m_\alpha ^2+3   m_\beta ^2  )  {\rm{B_{1}}}  (   m_\beta ^2,  m_{N_1} ^2,  m_{W_R} ^2  )\nonumber \\
 && 
-2   m_{W_R} ^4  {\rm{C_{0}}}  (   m_\alpha ^2,  m_{H_3} ^2,  m_\beta ^2,  m_{N_1} ^2,  m_{W_R} ^2,  m_{W_R} ^2  )
+ 
 2   m_{W_R} ^2  {\rm{C_{00}}}  (   m_\alpha ^2,  m_{H_3} ^2,  m_\beta ^2,  m_{N_1} ^2,  m_{W_R} ^2,  m_{W_R} ^2  )\nonumber \\
 &&+ 4 (   m_{H_3} ^2+2   m_\alpha ^2-   m_\beta ^2  )  {\rm{C_{001}}}  (   m_\alpha ^2,  m_{H_3} ^2,  m_\beta ^2,  m_{N_1} ^2,  m_{W_R} ^2,  m_{W_R} ^2  ) \nonumber \\
&&  +  4(   m_\alpha ^2-   m_{H_3} ^2  )  {\rm{C_{002}}}  (   m_\alpha ^2,  m_{H_3} ^2,  m_\beta ^2,  m_{N_1} ^2,  m_{W_R} ^2,  m_{W_R} ^2  )
 -6   m_{W_R} ^4  {\rm{C_{1}}}  (   m_\alpha ^2,  m_{H_3} ^2,  m_\beta ^2,  m_{N_1} ^2,  m_{W_R} ^2,  m_{W_R} ^2  ) \nonumber \\
&& +  m_{W_R} ^2 (
 3   m_\alpha ^2  +  m_\beta ^2   -  m_{H_3} ^2   )  {\rm{C_{11}}}  (   m_\alpha ^2,  m_{H_3} ^2,  m_\beta ^2,  m_{N_1} ^2,  m_{W_R} ^2,  m_{W_R} ^2  )\nonumber \\
 && + 2 m_\alpha ^2 ( m_{H_3} ^2  +  m_\alpha ^2-  m_\beta ^2  )  {\rm{C_{111}}}  (   m_\alpha ^2,  m_{H_3} ^2,  m_\beta ^2,  m_{N_1} ^2,  m_{W_R} ^2,  m_{W_R} ^2  )\nonumber \\
 &&- ( (m_{H_3} ^2- m_\beta ^2)^2 + m_\alpha^2 (2 m_{H_3} ^2 - 3   m_\alpha ^2 + 2 m_\beta ^2 ) )
 {\rm{C_{112}}}  (   m_\alpha ^2,  m_{H_3} ^2,  m_\beta ^2,  m_{N_1} ^2,  m_{W_R} ^2,  m_{W_R} ^2  )\nonumber \\
 &&+ m_{W_R} ^2 (
   m_\alpha ^2   +3   m_\beta ^2  -  m_{H_3} ^2     )  {\rm{C_{12}}}  (   m_\alpha ^2,  m_{H_3} ^2,  m_\beta ^2,  m_{N_1} ^2,  m_{W_R} ^2,  m_{W_R} ^2  )\nonumber \\
 &&+(( m_{H_3} ^2-m_\alpha ^2)^2-m_\beta ^4)
 {\rm{C_{122}}}  (   m_\alpha ^2,  m_{H_3} ^2,  m_\beta ^2,  m_{N_1} ^2,  m_{W_R} ^2,  m_{W_R} ^2  )\nonumber \\
 &&+ m_{W_R} ^2 (
  m_{H_3} ^2   -  m_\alpha ^2  +3   m_\beta ^2     )  {\rm{C_{2}}}  (   m_\alpha ^2,  m_{H_3} ^2,  m_\beta ^2,  m_{N_1} ^2,  m_{W_R} ^2,  m_{W_R} ^2  )\nonumber \\
 && + (  m_{H_3}^2 -  m_\alpha^2 )  (  m_{H_3} ^2-  m_\alpha ^2+  m_\beta ^2  )  {\rm{C_{22}}}  (   m_\alpha ^2,  m_{H_3} ^2,  m_\beta ^2,  m_{N_1} ^2,  m_{W_R} ^2,  m_{W_R} ^2  ) \,.
\end{eqnarray}
\end{widetext}
The corresponding couplings $c_{\alpha\beta}^{(L),6}$ and $c_{\alpha\beta}^{(R),6}$ for the diagram with the RHN $N_2$ can be obtained by substituting \(m_{N_1}\) with \(m_{N_2}\), and replacing \(\sin\theta\) with \(-\sin\theta\) in the prefactor.

\subsection{LFC couplings}

For the LFC couplings, the effective coupling $g_\ell$ can be decomposed as
\begin{equation}
\label{eqn:gl}
g_\ell = \sum_{i=1}^{6} \left( c_{\ell}^{(L),i} + c_{\ell}^{(R),i} \right) \,.
\end{equation}
Let us first consider the coupling of $H_3$ with $\ell_\beta$. For the first diagram in Fig.~\ref{fig:diagram}, the charged leptons running in the loop can be of either flavor $\alpha$ or $\beta$. The corresponding couplings are, respectively:
\begin{widetext}
\begin{eqnarray}
\label{eqn:cbeta:L1}
&&\left(\frac{m_\beta\sin\theta \cos\theta}{32 \sqrt{2} \pi ^2   v_R ^3}\right)^{-1} c_\beta^{(L),1} \ {\text{=}} \ \nonumber \\
&& 8  \sin 2   \theta \,  m_{H_R^{\pm\pm}} ^2 (  m_{N_1} -  m_{N_2} )^2   {\rm{C_1}}  (  m_{H_3} ^2,  m_\beta ^2,  m_\beta ^2,  m_{H_R^{\pm\pm}} ^2,  m_{H_R^{\pm\pm}} ^2,  m_\alpha ^2  )   \,, \\
&&\left(\frac{m_\beta\sin\theta \cos\theta}{32 \sqrt{2} \pi ^2   v_R ^3}\right)^{-1} c_\beta^{(R),1} \ {\text{=}}  \ \nonumber \\
&& -8 \sin 2   \theta\,  m_{H_R^{\pm\pm}} ^2 (  m_{N_1} -  m_{N_2} )^2   \left[ {\rm{C_0}}  (  m_{H_3} ^2,  m_\beta ^2,  m_\beta ^2,  m_{H_R^{\pm\pm}} ^2,  m_{H_R^{\pm\pm}} ^2,  m_\alpha ^2  ) \right. \nonumber \\
&& \left. + {\rm{C_1}}  (  m_{H_3} ^2,  m_\beta ^2,  m_\beta ^2,  m_{H_R^{\pm\pm}} ^2,  m_{H_R^{\pm\pm}} ^2,  m_\alpha ^2  )+ {\rm{C_2}}  (  m_{H_3} ^2,  m_\beta ^2,  m_\beta ^2,  m_{H_R^{\pm\pm}} ^2,  m_{H_R^{\pm\pm}} ^2,  m_\alpha ^2  ) \right]   \,,    \\
 &&\left(\frac{m_\beta\sin\theta \cos\theta}{32 \sqrt{2} \pi ^2   v_R ^3}\right)^{-1}c_\beta^{(L),2} \ {\text{=}}   \ \nonumber \\
 &&  16 \tan \theta \, m_{H_R^{\pm\pm}} ^2 (  m_{N_1}  \sin\theta+  m_{N_2}  \cos\theta \cot \theta)^2     {\rm{C_1}}  (  m_{H_3} ^2,  m_\beta ^2,  m_\beta ^2,  m_{H_R^{\pm\pm}} ^2,  m_{H_R^{\pm\pm}} ^2,  m_\beta ^2  )     \,, \\
 &&\left(\frac{m_\beta\sin\theta \cos\theta}{32 \sqrt{2} \pi ^2   v_R ^3}\right)^{-1} c_\beta^{(R),2} \ {\text{=}}  \ \nonumber \\
 &&    -16 \sin ^2\theta \tan \theta  \,  m_{H_R^{\pm\pm}} ^2  (  m_{N_1} +  m_{N_2}  \cot ^2\theta  )^2 \left[ {\rm{C_0}}  (  m_{H_3} ^2,  m_\beta ^2,  m_\beta ^2,  m_{H_R^{\pm\pm}} ^2,  m_{H_R^{\pm\pm}} ^2,  m_\beta ^2  ) \right. \nonumber \\
 && \left. +  {\rm{C_1}}  (  m_{H_3} ^2,  m_\beta ^2,  m_\beta ^2,  m_{H_R^{\pm\pm}} ^2,  m_{H_R^{\pm\pm}} ^2,  m_\beta ^2  )+  {\rm{C_2}}  (  m_{H_3} ^2,  m_\beta ^2,  m_\beta ^2,  m_{H_R^{\pm\pm}} ^2,  m_{H_R^{\pm\pm}} ^2,  m_\beta ^2  )  \right] \,.
\end{eqnarray}

For the second diagram in Fig.~\ref{fig:diagram} with the RHN $N_1$, the couplings are:
\begin{eqnarray}
&&\left(\frac{m_\beta\sin\theta \cos\theta}{32 \sqrt{2} \pi ^2   v_R ^3}\right)^{-1} c_\beta^{(L),3} \ \simeq \  \nonumber \\
&&    m_{N_1} ^2\tan \theta \left\{ 6      {\rm{B_{0}}}  (  m_\beta ^2,  m_{N_1} ^2,  m_{W_R} ^2  )+4 {\rm{B_{1}}}  (  m_\beta ^2,  m_{N_1} ^2,  m_{W_R} ^2  ) \right. \nonumber \\
&&+  2 ( 
 m_{N_1} ^2-2   m_{W_R} ^2  )  {\rm{C_0}}  (  m_{H_3} ^2,  m_\beta ^2,  m_\beta ^2,  m_{N_1} ^2,  m_{N_1} ^2,  m_{W_R} ^2  )-  4{\rm{C_{00}}}  (  m_{H_3} ^2,  m_\beta ^2,  m_\beta ^2,  m_{N_1} ^2,  m_{N_1} ^2,  m_{W_R} ^2  )\nonumber \\
&&-  4 ( 
m_{N_1} ^2+2   m_{W_R} ^2  )  {\rm{C_1}}  (  m_{H_3} ^2,  m_\beta ^2,  m_\beta ^2,  m_{N_1} ^2,  m_{N_1} ^2,  m_{W_R} ^2  )-4   m_{H_3} ^2    {\rm{C_{11}}}  (  m_{H_3} ^2,  m_\beta ^2,  m_\beta ^2,  m_{N_1} ^2,  m_{N_1} ^2,  m_{W_R} ^2  )\nonumber \\
&& \left. -2   m_{H_3} ^2    {\rm{C_{12}}}  (  m_{H_3} ^2,  m_\beta ^2,  m_\beta ^2,  m_{N_1} ^2,  m_{N_1} ^2,  m_{W_R} ^2  )+4   m_\beta ^2    {\rm{C_2}}  (  m_{H_3} ^2,  m_\beta ^2,  m_\beta ^2,  m_{N_1} ^2,  m_{N_1} ^2,  m_{W_R} ^2  )   \right\} \,, \\
&&\left(\frac{m_\beta\sin\theta \cos\theta}{32 \sqrt{2} \pi ^2   v_R ^3}\right)^{-1} c_\beta^{(R),3} \ \simeq \ \nonumber \\
&&    m_{N_1} ^2 \tan \theta \left\{ 2    {\rm{B_{0}}}  (  m_\beta ^2,  m_{N_1} ^2,  m_{W_R} ^2  )
+2  (  
3   m_{N_1} ^2+2   m_{W_R} ^2  )  {\rm{C_0}}  (  m_{H_3} ^2,  m_\beta ^2,  m_\beta ^2,  m_{N_1} ^2,  m_{N_1} ^2,  m_{W_R} ^2  ) \right. \nonumber \\
&&+4     {\rm{C_{00}}}  (  m_{H_3} ^2,  m_\beta ^2,  m_\beta ^2,  m_{N_1} ^2,  m_{N_1} ^2,  m_{W_R} ^2  )
+4     ( 
  m_{N_1} ^2+2   m_{W_R} ^2  )  {\rm{C_1}}  (  m_{H_3} ^2,  m_\beta ^2,  m_\beta ^2,  m_{N_1} ^2,  m_{N_1} ^2,  m_{W_R} ^2  \nonumber \\
&&+4   m_{H_3} ^2    {\rm{C_{11}}}  (  m_{H_3} ^2,  m_\beta ^2,  m_\beta ^2,  m_{N_1} ^2,  m_{N_1} ^2,  m_{W_R} ^2  )+6   m_{H_3} ^2    {\rm{C_{12}}}  (  m_{H_3} ^2,  m_\beta ^2,  m_\beta ^2,  m_{N_1} ^2,  m_{N_1} ^2,  m_{W_R} ^2  )\nonumber \\
&& \left. +4    ( 
m_{N_1} ^2+2   m_{W_R} ^2  )  {\rm{C_2}}  (  m_{H_3} ^2,  m_\beta ^2,  m_\beta ^2,  m_{N_1} ^2,  m_{N_1} ^2,  m_{W_R} ^2  )+2   m_{H_3} ^2     {\rm{C_{22}}}  (  m_{H_3} ^2,  m_\beta ^2,  m_\beta ^2,  m_{N_1} ^2,  m_{N_1} ^2,  m_{W_R} ^2  )   \right\} \,. \nonumber \\
\end{eqnarray}
The corresponding couplings $c_{\beta}^{(L),4}$ and $c_{\beta}^{(R),4}$ for the diagram with the RHN $N_2$ can be obtained by substituting \(m_{N_1}\) with \(m_{N_2}\), and replacing \(\tan\theta\) with \(\cot\theta\) in the prefactor on the R.H.S..


For the last diagram in Fig.~\ref{fig:diagram} with the RHN $N_1$, the couplings are:
\begin{eqnarray}
 &&\left(\frac{m_\beta\sin\theta \cos\theta}{32 \sqrt{2} \pi ^2   v_R ^3}\right)^{-1} c_{\beta}^{(L),5} \ \simeq \  \nonumber \\
 &&   \tan \theta \Big\{    ( 
 -2   m_{N_1} ^2+2   m_{W_R} ^2  )  {\rm{B_{0}}}  (  m_\beta ^2,  m_{N_1} ^2,  m_{W_R} ^2  )-2   {\rm{B_{00}}}  (  m_\beta ^2,  m_{N_1} ^2,  m_{W_R} ^2  )  \nonumber \\
    &&+2   (  m_{W_R} ^2-  m_{N_1} ^2  )   {\rm{B_{1}}}  (  m_\beta ^2,  m_{N_1} ^2,  m_{W_R} ^2  )-  m_{H_3} ^2 {\rm{B_{11}}}  (  m_\beta ^2,  m_{N_1} ^2,  m_{W_R} ^2  )\nonumber \\
    &&+   
    2   m_{W_R} ^4   {\rm{C_0}}  (  m_\beta ^2,  m_{H_3} ^2,  m_\beta ^2,  m_{N_1} ^2,  m_{W_R} ^2,  m_{W_R} ^2  )
    -2   m_{W_R} ^2   {\rm{C_{00}}}  (  m_\beta ^2,  m_{H_3} ^2,  m_\beta ^2,  m_{N_1} ^2,  m_{W_R} ^2,  m_{W_R} ^2  )\nonumber \\
    &&+4   m_\beta ^2   {\rm{C_{001}}}  (  m_\beta ^2,  m_{H_3} ^2,  m_\beta ^2,  m_{N_1} ^2,  m_{W_R} ^2,  m_{W_R} ^2  )+  (4   m_\beta ^2-4   m_{H_3} ^2  )  {\rm{C_{002}}}  (  m_\beta ^2,  m_{H_3} ^2,  m_\beta ^2,  m_{N_1} ^2,  m_{W_R} ^2,  m_{W_R} ^2  )\nonumber \\
    &&-   
     m_{W_R} ^2(  m_{H_3} ^2 + 2   m_\beta ^2   )  {\rm{C_1}}  (  m_\beta ^2,  m_{H_3} ^2,  m_\beta ^2,  m_{N_1} ^2,  m_{W_R} ^2,  m_{W_R} ^2  )-  m_{H_3} ^2   (  m_{H_3} ^2-2   m_\beta ^2  )  {\rm{C_{222}}}  (  m_\beta ^2,  m_{H_3} ^2,  m_\beta ^2,  m_{N_1} ^2,  m_{W_R} ^2,  m_{W_R} ^2  ) \nonumber \\
    && -  m_{H_3} ^2   m_\beta ^2 \left[  {\rm{C_{11}}}  (  m_\beta ^2,  m_{H_3} ^2,  m_\beta ^2,  m_{N_1} ^2,  m_{W_R} ^2,  m_{W_R} ^2  )+2     {\rm{C_{112}}}  (  m_\beta ^2,  m_{H_3} ^2,  m_\beta ^2,  m_{N_1} ^2,  m_{W_R} ^2,  m_{W_R} ^2  ) \right] \nonumber \\
   && +    
    m_{W_R} ^2  ( m_{H_3} ^2  -4   m_\beta ^2 )  \left[ {\rm{C_{12}}}  (  m_\beta ^2,  m_{H_3} ^2,  m_\beta ^2,  m_{N_1} ^2,  m_{W_R} ^2,  m_{W_R} ^2  )+{\rm{C_{22}}}  (  m_\beta ^2,  m_{H_3} ^2,  m_\beta ^2,  m_{N_1} ^2,  m_{W_R} ^2,  m_{W_R} ^2  ) \right] \nonumber \\
    &&+  m_{H_3} ^4  {\rm{C_{122}}}  (  m_\beta ^2,  m_{H_3} ^2,  m_\beta ^2,  m_{N_1} ^2,  m_{W_R} ^2,  m_{W_R} ^2  )
    +   
    6   m_{W_R} ^4    {\rm{C_2}}  (  m_\beta ^2,  m_{H_3} ^2,  m_\beta ^2,  m_{N_1} ^2,  m_{W_R} ^2,  m_{W_R} ^2  )
 \Big\} \,, \\
 &&\left(\frac{m_\beta\sin\theta \cos\theta}{32 \sqrt{2} \pi ^2   v_R ^3}\right)^{-1} c_{\beta}^{(R),5} \ \simeq \   \nonumber \\
 &&  \tan \theta \Big\{  -2  {A_0}  (  m_{W_R} ^2  )   + (
 2   m_{W_R} ^2 -2   m_{N_1} ^2 )  {\rm{B_{0}}}  (  m_\beta ^2,  m_{N_1} ^2,  m_{W_R} ^2  ) \nonumber \\
 && +2   {\rm{B_{00}}}  (  m_\beta ^2,  m_{N_1} ^2,  m_{W_R} ^2  )-  (  m_{H_3} ^2+2   m_\beta ^2  )  {\rm{B_{1}}}  (  m_\beta ^2,  m_{N_1} ^2,  m_{W_R} ^2  )\nonumber \\
 && 
 -   
   2 m_{W_R} ^2 {\rm{C_{00}}}  (  m_\beta ^2,  m_{H_3} ^2,  m_\beta ^2,  m_{N_1} ^2,  m_{W_R} ^2,  m_{W_R} ^2 ) - m_{W_R}^2 (
 m_{H_3} ^2   +2   m_\beta ^2   )  {\rm{C_2}}  (  m_\beta ^2,  m_{H_3} ^2,  m_\beta ^2,  m_{N_1} ^2,  m_{W_R} ^2,  m_{W_R} ^2  ) \nonumber \\
 &&- 4 (   m_{H_3} ^2 + m_\beta ^2  )   {\rm{C_{001}}}  (  m_\beta ^2,  m_{H_3} ^2,  m_\beta ^2,  m_{N_1} ^2,  m_{W_R} ^2,  m_{W_R} ^2  )-  4 ( m_\beta ^2 - m_{H_3} ^2  )   {\rm{C_{002}}}  (  m_\beta ^2,  m_{H_3} ^2,  m_\beta ^2,  m_{N_1} ^2,  m_{W_R} ^2,  m_{W_R} ^2  )\nonumber \\
 &&  + 
2   m_{W_R} ^4  \left[  {\rm{C_0}}  (  m_\beta ^2,  m_{H_3} ^2,  m_\beta ^2,  m_{N_1} ^2,  m_{W_R} ^2,  m_{W_R} ^2  ) + 3   
{\rm{C_1}}  (  m_\beta ^2,  m_{H_3} ^2,  m_\beta ^2,  m_{N_1} ^2,  m_{W_R} ^2,  m_{W_R} ^2  ) \right]
 \nonumber \\
 &&-2   m_{H_3} ^2   m_\beta ^2   {\rm{C_{111}}}  (   m_\beta ^2,  m_{H_3} ^2,  m_\beta ^2,  m_{N_1} ^2,  m_{W_R} ^2,  m_{W_R} ^2  )+  m_{H_3} ^4   {\rm{C_{112}}}  (  m_\beta ^2,  m_{H_3} ^2,  m_\beta ^2,  m_{N_1} ^2,  m_{W_R} ^2,  m_{W_R} ^2  )\nonumber \\
 &&-   m_{W_R} ^2 (
 4   m_\beta ^2   -  m_{H_3} ^2  ) \left[ {\rm{C_{11}}}  (  m_\beta ^2,  m_{H_3} ^2,  m_\beta ^2,  m_{N_1} ^2,  m_{W_R} ^2,  m_{W_R} ^2  )+{\rm{C_{12}}}  (  m_\beta ^2,  m_{H_3} ^2,  m_\beta ^2,  m_{N_1} ^2,  m_{W_R} ^2,  m_{W_R} ^2  ) \right] \nonumber \\
 && - m_{H_3}^2  ( m_{H_3} ^2 - 2 m_\beta ^2  )  {\rm{C_{122}}}  (  m_\beta ^2,  m_{H_3} ^2,  m_\beta ^2,  m_{N_1} ^2,  m_{W_R} ^2,  m_{W_R} ^2  )\nonumber \\
 &&-  m_{H_3} ^2 (  m_{H_3}^2 -  m_\beta^2 )    {\rm{C_{22}}}  (  m_\beta ^2,  m_{H_3} ^2,  m_\beta ^2,  m_{N_1} ^2,  m_{W_R} ^2,  m_{W_R} ^2  )  \Big\} \,.
\end{eqnarray}
The corresponding couplings $c_{\beta}^{(L),6}$ and $c_{\beta}^{(R),6}$ for the diagram with the RHN $N_2$ can be obtained by substituting \(m_{N_1}\) with \(m_{N_2}\), and replacing \(\tan\theta\) with \(\cot\theta\) in the prefactor on the R.H.S..


For the coupling of $H_3$ with the charged lepton $\ell_\alpha$, the coefficients for the first diagram in Fig.~\ref{fig:diagram} are a little bit different from the case of $\ell_\beta$:
\begin{eqnarray}
 &&\left(\frac{m_\alpha\sin\theta \cos\theta}{32 \sqrt{2} \pi ^2   v_R ^3}\right)^{-1} c_{\alpha}^{(L),1} \ {\text{=}}  \ \nonumber \\
 &&  4 \sec \theta \csc \theta \,  m_{H_R^{\pm\pm}} ^2  ((  m_{N_1} -  m_{N_2} ) \cos 2   \theta
 +  m_{N_1} +  m_{N_2} )^2  {\rm{C_1}}  (  m_{H_3} ^2,  m_\alpha ^2,  m_\alpha ^2,  m_{H_R^{\pm\pm}} ^2,  m_{H_R^{\pm\pm}} ^2,  m_\alpha ^2  )   \,, \\
 &&\left(\frac{m_\alpha\sin\theta \cos\theta}{32 \sqrt{2} \pi ^2   v_R ^3}\right)^{-1} c_{\alpha}^{(R),1} \ {\text{=}} \  \nonumber \\
 && -4 \sec \theta \csc \theta \,  m_{H_R^{\pm\pm}} ^2  ((  m_{N_1} -  m_{N_2} ) \cos 2   \theta +  m_{N_1} +  m_{N_2} )^2  \Big[ {\rm{C_0}}  (  m_{H_3} ^2,  m_\alpha ^2,  m_\alpha ^2, m_{H_R^{\pm\pm}} ^2,  m_{H_R^{\pm\pm}} ^2,  m_\alpha ^2  ) \nonumber \\
 && + {\rm{C_1}}  (  m_{H_3} ^2,  m_\alpha ^2,  m_\alpha ^2,  m_{H_R^{\pm\pm}} ^2,  m_{H_R^{\pm\pm}} ^2,  m_\alpha ^2  )+  {\rm{C_2}}  (  m_{H_3} ^2,  m_\alpha ^2,  m_\alpha ^2,  m_{H_R^{\pm\pm}} ^2,  m_{H_R^{\pm\pm}} ^2,  m_\alpha ^2  ) \Big]  \,, \\
 && \left(\frac{m_\alpha\sin\theta \cos\theta}{32 \sqrt{2} \pi ^2   v_R ^3}\right)^{-1}c_{\alpha}^{(L),2} \ {\text{=}} \ \nonumber \\
 && 8 \sin 2   \theta \, m_{H_R^{\pm\pm}} ^2 (  m_{N_1} -  m_{N_2} )^2   {\rm{C_1}}  (  m_{H_3} ^2,  m_\alpha ^2,  m_\alpha ^2,  m_{H_R^{\pm\pm}} ^2,  m_{H_R^{\pm\pm}} ^2,  m_\beta ^2  )  \,, \\
\label{eqn:calpha:R2}
 &&\left(\frac{m_\alpha\sin\theta \cos\theta}{32 \sqrt{2} \pi ^2   v_R ^3}\right)^{-1}c_{\alpha}^{(R),2} \ {\text{=}}  \ \nonumber \\
 && -8 \sin 2   \theta \,  m_{H_R^{\pm\pm}} ^2 (  m_{N_1} -  m_{N_2} )^2  \Big[ {\rm{C_0}}  (  m_{H_3} ^2,  m_\alpha ^2,  m_\alpha ^2,  m_{H_R^{\pm\pm}} ^2,  m_{H_R^{\pm\pm}} ^2,  m_\beta ^2  ) \nonumber \\
 && +  {\rm{C_1}}  (  m_{H_3} ^2,  m_\alpha ^2,  m_\alpha ^2,  m_{H_R^{\pm\pm}} ^2,  m_{H_R^{\pm\pm}} ^2,  m_\beta ^2  )+ {\rm{C_2}}  (  m_{H_3} ^2,  m_\alpha ^2,  m_\alpha ^2,  m_{H_R^{\pm\pm}} ^2,  m_{H_R^{\pm\pm}} ^2,  m_\beta ^2  ) \Big]  \,.
\end{eqnarray}
\end{widetext}
For the second and third diagram with the RHN $N_{1}$ running in the loop, the corresponding couplings $c_{\alpha}^{(L,R),3}$ and $c_{\alpha}^{(L,R),5}$ can be obtained from $c_{\beta}^{(L,R),3}$ and $c_{\beta}^{(L,R),5}$ by substituting \(m_{\beta}\) with \(m_{\alpha}\), and replacing \(\tan\theta\) with \(\cot\theta\) in the prefactor on the R.H.S., respectively. If it is the RHN $N_2$ in the loop, to obtain the couplings $c_{\alpha}^{(L,R),4}$ and $c_{\alpha}^{(L,R),6}$, we need to substitute \(m_{N_1}\) with \(m_{N_2}\), and replace \(m_{\beta}\) with \(m_{\alpha}\) in $c_{\beta}^{(L,R),4}$ and $c_{\beta}^{(L,R),6}$, respectively.

\section{Comparison with couplings of ALPs}
\label{app:comparison}

For convenience, the LFV couplings of ALP can be written as
\begin{equation}
{\cal L} = - i g_{a \, \alpha\beta} a \bar{\ell}_\alpha \gamma_5 \ell_\beta ~+~ {\rm H.c.}  \,.
\end{equation}
In the limit of hierarchical structure of charged lepton masses $m_{\ell_\alpha} \ll m_{\ell_\beta}$, the squared amplitudes for the LFV processes $\ell_\beta \to \ell_\alpha + a$ and $a \to \ell_\alpha + \ell_\beta$ are both proportional to the factor of $(m_{\ell_\beta}^2 - m_a^2)$,
which has the same form as that in Eqs.~(\ref{eqn:width:LFV}) and (\ref{eqn:width:LFV2}) for the CP-even $H_3$ in the limit of $m_{\ell_\alpha} \to 0$. Similarly, for the lepton-proton bremsstrahlung $\ell_\alpha + p \to \ell_\beta + p + H_3/a$  and semi-Compton $\ell_\alpha + \gamma \to \ell_\beta +H_3/a$ processes which are relevant to the supernova constraints, the cross sections are the same for scalars and pseudoscalars in the limit of $m_{\ell_\alpha} \to 0$.
See Ref.~\cite{Qiang:2026abc} for more details.

The LFC couplings of an ALP can be written as
\begin{equation}
{\cal L} = - i g_{a\ell} a \bar\ell \gamma_5 \ell \,.
\end{equation}
If the charged lepton $\ell$ involved is relativistic, the squared amplitudes for scalars and pseudoscalars are both proportional to the squared (pseudo)scalar mass. Therefore, the ALP constraints can be directly converted to the limits on $H_3$ for such cases.

The coupling of ALPs to photons can be written in the form of
\begin{equation}
{\cal L} = - \frac{1}{4} g_{a\gamma\gamma} a F_{\mu\nu} \tilde{F}^{\mu\nu} \,,
\end{equation}
with $\tilde{F}^{\mu\nu} = \epsilon^{\mu\nu\rho\sigma}F_{\rho\sigma}/2$ the dual of the electromagnetic field. For both the processes of $H_3/a \leftrightarrow \gamma\gamma$ and $\gamma + f \to f + H_3/a$ (with the fermion $f$ either relativistic or non-relativistic), the squared amplitudes are the same for scalars and pseudoscalars in the limit of massless photon~\cite{Qiang:2026abc}. The Primakoff-like process $\gamma + f \to f + H_3/a$ is a good approximation for the production of $H_3/a$ in the proton and electron beam-dump experiments. In compact stars such as supernovae and neutron stars, the Primakoff and photon coalescence processes are the dominant production channels of $H_3$ and $a$, and $H_3/a \to \gamma\gamma$ are relevant to their decays.

\section{More details on the constraints}
\label{app:limit}

All the predominant current constraints on the LFV couplings $g_{\alpha\beta}$ and the LFC coupling $g_e$ of $H_3$ from the laboratory experiments and astrophysical observations and the leading future prospects are collected in Table~\ref{tab:LFV}. The leading constraints on the coupling of $H_3$ with photons and future prospects are presented in Table~\ref{tab:gauge}. In both the two tables, the corresponding constraints on the $v_R$ scale are listed in the last columns. Here are more details for some of the constraints.

\begin{table*}[t!]
  \centering
  \caption[]{Leading current limits and future prospects of the LFV couplings $g_{\alpha\beta}$ and the LFC coupling $g_e$ of $H_3$ from the laboratory experiments and astrophysical observations. The corresponding limits or prospects of $v_R$ are collected in the last column. See Fig.~\ref{fig:LFV} and text for more details.   }
  \label{tab:LFV}
  \begin{tabular}{|c|c|c|c|c|c|c|}
  \hline\hline
  & coupling & experiment & {process} & mass [MeV] & {limit} & {$v_R$ [GeV]} \\ \hline 

  \parbox[t]{2mm}{\multirow{9}{*}{\rotatebox[origin=c]{90}{laboratory}}} & \multirow{4}{*}{$g_{e\mu}$} & TWIST~\cite{TWIST:2014ymv} & $ \mu \to e + {\rm inv}$  & $\lesssim 80$ & $ {{\rm BR}\sim10^{-5}}$ & $5\times10^8$\\ \cline{3-7}

  & & PIENU~\cite{PIENU:2020loi} & $ \mu\to e+ {\rm inv}$  & $(47.8,\,95.1)$ & $ {\rm BR} \sim10^{-5}$ &$7\times10^7$ \\ \cline{3-7}

  & & Jodidio et al~\cite{Jodidio:1986mz} & $ \mu\to e+ {\rm inv}$  & $\lesssim10$ & $ {\rm BR} \sim10^{-6}$ &$2\times10^8$  \\ \cline{3-7}

  & & Mu3e~\cite{Banerjee:2022nbr,Perrevoort:2018ttp} & $ \mu\to e+{\rm inv}$  & $\lesssim 100$ & ${\rm BR}\sim10^{-8}$ & $5\times10^9$\\ \cline{2-7}

  & \multirow{2}{*}{$g_{e\tau}$} & Belle~\cite{Belle:2025bpu} & $ \tau\to e+ {\rm inv}$  & $<1600$ & ${\rm BR} \sim 10^{-4}$ & $3\times10^5$ \\ \cline{3-7}

  & & Belle II~\cite{DeLaCruz-Burelo:2020ozf} & $ \tau\to e+{\rm inv}$  & $<1600$ & ${\rm BR} \sim10^{-6}$ & $1\times10^7$  \\ \cline{2-7}

  & $g_{\mu\tau}$ & Belle~\cite{Belle:2025bpu} & $ \tau\to \mu+ {\rm inv}$  & $<1600$ & ${\rm BR}\sim 10^{-4}$ & $4\times10^5$ \\ \cline{2-7}

  & \multirow{2}{*}{$g_{\ell\tau}$} & CHARM~\cite{Ema:2025bww} & $ \tau \to \ell + a$  & $\lesssim 1000$ & $  f_a\sim 10^8~\rm{GeV}$ & $5\times10^5$\\ \cline{3-7}

  & & SHiP~\cite{Ema:2025bww}  & $ \tau \to \ell+a$  & $\lesssim 1000$ & $f_a \sim 10^{9}$ GeV &$2\times10^7$ \\ \hline

  \parbox[t]{2mm}{\multirow{5}{*}{\rotatebox[origin=c]{90}{astro.}}} & \multirow{4}{*}{$g_{e\mu}$} & \multirow{2}{*}{SN1987A~\cite{Li:2025beu}} & $\mu \to e + a$, $e+\mu \to a$ & \multirow{2}{*}{$\lesssim 280$} & \multirow{2}{*}{$g_{aeu} \sim 10^{-9}$} & \multirow{2}{*}{$2\times10^7$}
  \\ 
  &&& $\ell_\alpha + p \to \ell_\beta + p + a$ && & \\ \cline{3-7}

  & & \multirow{2}{*}{LESNe~\cite{Huang:2025rmy,Huang:2025xvo}} & $\mu \to e + a$, $e + \mu \to a$ & \multirow{2}{*}{$\lesssim500$} & \multirow{2}{*}{$g_{aeu} \sim 10^{-11}$} & \multirow{2}{*}{$4\times10^8$} \\
  && & $e + \gamma \to \mu + a$ && & \\
  \cline{2-7}
  & $g_{e}$ & LESNe~\cite{Fiorillo:2025sln} & $e + \gamma \to e + a$, $e^+ + e^- \to a$  & $\lesssim 200$ & $ g_{ae} \sim10^{-11}$ &$1\times10^6$ \\ \hline\hline
  \end{tabular}
\end{table*}

\begin{table*}[t!]
  \centering
  \caption[]{The same as Fig.~\ref{tab:LFV}, but for the astrophysical constraints on the coupling of $H_3$ with photons. See Fig.~\ref{fig:LFV} and text for more details.
  }
  \label{tab:gauge}
  \begin{tabular}{|c|c|c|c|c|}
  \hline\hline
  \multirow{2}{*}{experiment} & \multirow{2}{*}{process} & \multirow{2}{*}{mass [MeV]} & limits on $g_{a\gamma\gamma}$ & \multirow{2}{*}{$v_R$ [GeV]} \\
  &  & & [GeV$^{-1}$] & \\ \hline

SN1987A~\cite{Muller:2023vjm} & \multirow{7}{*}{$\gamma + {\cal N} \to {\cal N} + a$, $\gamma + \gamma \to a$} &  $\lesssim 300$ & $\sim 10^{-12}$ &   $2\times10^9$\\ \cline{1-1} \cline{3-5}

SN2023ixf~\cite{Muller:2023pip} & & $\lesssim 4$ & $\sim 10^{-11.5}$ & $7\times10^8$ \\ \cline{1-1} \cline{3-5}

type Ic supernovae~\cite{Candon:2025ypl} & & $\lesssim 400$ & $\sim10^{-11}$ & $3\times10^8$ \\ \cline{1-1} \cline{3-5}

LESNe~\cite{Fiorillo:2025yzf} & & $\lesssim 550$ & $\sim 10^{-10}$ &  $2\times10^7$ \\  \cline{1-1} \cline{3-5}

GW170817~\cite{Diamond:2023cto} & & $\lesssim 400$  & $\sim 10^{-10}$ & $5\times10^7$ \\ \cline{1-1} \cline{3-5}

Betelgeuse~\cite{Jaeckel:2017tud} & & $\lesssim 100$ & $\sim 10^{-14}$ & $6\times10^{11}$\\  \cline{1-1} \cline{3-5}

NS $\gamma$-rays~\cite{Dev:2023hax} & & $\lesssim660$ & $\sim 10^{-12}$ &$3\times10^9$ \\ \hline\hline
\end{tabular}
\end{table*}

The constraints of beam dump experiments on ALP in Ref.~\cite{Ema:2025bww} can be re-interpreted and applied to the light $H_3$ in the minimal LRSM. The light $H_3$ and $a$ are to some extent different, in particular their decay modes. The ALP $a$ is assumed to decay mostly into dileptons $ee$, $e\mu$ and $\mu\mu$ in Ref.~\cite{Ema:2025bww}, whereas $H_3$ decays predominantly into diphoton for some of the mass ranges (cf. the left panel of Fig.~\ref{fig:BR}). However, all these decay products are visible particles, and can be easily identified at the detectors. For concreteness, we adopt the anarchical LFV case with universal ALP couplings to the charged leptons in Fig. 6 of Ref.~\cite{Ema:2025bww}, and recast the current limits from CHARM and the prospect of SHiP in the following way.
\begin{itemize}
    \item The lower boundaries of these constraints are determined by the production of $a$ via the process of $\tau \to \ell + a$. The corresponding lower boundaries on the $f_a$ parameter for ALP can be directly converted to the LFV coupling $g_{\alpha\beta}$ via the relation of $g_{\tau\ell} = m_\tau / f_a$.
\item The upper boundaries of these constraints are mainly dictated by the decay of ALP. Therefore, setting $H_3$ and $a$ have the same decay length, i.e. $\tau_{H_3} = \tau_a$ (with $\tau_a$ the proper lifetime of $a$), we can convert these upper boundaries onto the constraints on $m_{H_3}$ and $v_R$.
\end{itemize}

We take the energy deposition limit on $g_{ae}$ in the left panel of Fig.~8 of Ref.~\cite{Fiorillo:2025sln} to set limits on the light $H_3$. To this end, we adopt the following procedure for the re-interpretation.
\begin{itemize}
    \item The lower boundary is mainly determined by the production of $a$ in the supernova core via the $e + \gamma \to e + a$ and $e^+ + e^- \to a$ processes. The corresponding limits on $g_{ae}$ can be directly converted to the constraints on $v_R$ in the LRSM.

    \item
    The upper boundary for the energy deposition in the mantle of LESNe is determined mainly by ALP decay within the progenitor. In other words, to effectively deposit energy in the mantle, the lifetime of ALP is required to be within certain range. To set limits on $H_3$, we require the lifetime of $H_3$ to equal that of axion, i.e. $\tau_{H_3} = \tau_a$, with the latter one dominated by the decay $a \to e^+ e^-$.
\end{itemize}


\bibliography{ref_arXiv_v2}

@article{Dev:2017dui,
    author = "Dev, P. S. Bhupal and Mohapatra, Rabindra N. and Zhang, Yongchao",
    title = "{Long Lived Light Scalars as Probe of Low Scale Seesaw Models}",
    eprint = "1703.02471",
    archivePrefix = "arXiv",
    primaryClass = "hep-ph",
    reportNumber = "ULB-TH-17-05, UMD-PP-017-21",
    doi = "10.1016/j.nuclphysb.2017.07.021",
    journal = "Nucl. Phys. B",
    volume = "923",
    pages = "179--221",
    year = "2017"
}

@article{Dev:2017ftk,
    author = "Dev, P. S. Bhupal and Mohapatra, Rabindra N. and Zhang, Yongchao",
    title = "{Lepton Flavor Violation Induced by a Neutral Scalar at Future Lepton Colliders}",
    eprint = "1711.08430",
    archivePrefix = "arXiv",
    primaryClass = "hep-ph",
    doi = "10.1103/PhysRevLett.120.221804",
    journal = "Phys. Rev. Lett.",
    volume = "120",
    number = "22",
    pages = "221804",
    year = "2018"
}

@article{Huang:2025rmy,
    author = "Huang, Zi-Miao and Liu, Zuowei",
    title = "{Low-energy supernova constraints on lepton flavor violating axions}",
    eprint = "2506.16922",
    archivePrefix = "arXiv",
    primaryClass = "hep-ph",
    doi = "10.1007/JHEP10(2025)024",
    journal = "JHEP",
    volume = "10",
    pages = "024",
    year = "2025"
}

@article{Zhang:2023vva,
    author = "Zhang, Hong-Yi and Hagimoto, Ray and Long, Andrew J.",
    title = "{Neutron star cooling with lepton-flavor-violating axions}",
    eprint = "2309.03889",
    archivePrefix = "arXiv",
    primaryClass = "hep-ph",
    doi = "10.1103/PhysRevD.109.103005",
    journal = "Phys. Rev. D",
    volume = "109",
    number = "10",
    pages = "103005",
    year = "2024"
}

@article{Li:2025beu,
    author = "Li, Yonglin and Liu, Zuowei",
    title = "{Supernova constraints on lepton flavor violating axions}",
    eprint = "2501.12075",
    archivePrefix = "arXiv",
    primaryClass = "hep-ph",
    doi = "10.1103/dmnx-3t96",
    journal = "Phys. Rev. D",
    volume = "113",
    number = "5",
    pages = "055039",
    year = "2026"
}

@article{Calibbi:2020jvd,
    author = "Calibbi, Lorenzo and Redigolo, Diego and Ziegler, Robert and Zupan, Jure",
    title = "{Looking forward to lepton-flavor-violating ALPs}",
    eprint = "2006.04795",
    archivePrefix = "arXiv",
    primaryClass = "hep-ph",
    reportNumber = "P3H-20-024, TTP20-025",
    doi = "10.1007/JHEP09(2021)173",
    journal = "JHEP",
    volume = "09",
    pages = "173",
    year = "2021"
}

@article{Cui:2021dkr,
    author = "Cui, Chuan-Xin and Ishida, Hiroyuki and Matsuzaki, Shinya and Shigekami, Yoshihiro",
    title = "{Probing an intrinsically flavorful ALP via tau-lepton flavor physics}",
    eprint = "2110.11640",
    archivePrefix = "arXiv",
    primaryClass = "hep-ph",
    doi = "10.1103/PhysRevD.105.095033",
    journal = "Phys. Rev. D",
    volume = "105",
    number = "9",
    pages = "095033",
    year = "2022"
}

@article{Panci:2022wlc,
    author = "Panci, Paolo and Redigolo, Diego and Schwetz, Thomas and Ziegler, Robert",
    title = "{Axion dark matter from lepton flavor-violating decays}",
    eprint = "2209.03371",
    archivePrefix = "arXiv",
    primaryClass = "hep-ph",
    doi = "10.1016/j.physletb.2023.137919",
    journal = "Phys. Lett. B",
    volume = "841",
    pages = "137919",
    year = "2023"
}

@article{Cheung:2021mol,
    author = "Cheung, Kingman and Soffer, Abner and Wang, Zeren Simon and Wu, Yu-Heng",
    title = "{Probing charged lepton flavor violation with axion-like particles at Belle II}",
    eprint = "2108.11094",
    archivePrefix = "arXiv",
    primaryClass = "hep-ph",
    doi = "10.1007/JHEP11(2021)218",
    journal = "JHEP",
    volume = "11",
    pages = "218",
    year = "2021"
}

@article{Dev:2024ygx,
    author = "Dev, P. S. Bhupal and Kim, Doojin and Sathyan, Deepak and Sinha, Kuver and Zhang, Yongchao",
    title = "{New laboratory constraints on neutrinophilic mediators}",
    eprint = "2407.12738",
    archivePrefix = "arXiv",
    primaryClass = "hep-ph",
    reportNumber = "CETUP-2024-005",
    doi = "10.1016/j.physletb.2025.139765",
    journal = "Phys. Lett. B",
    volume = "868",
    pages = "139765",
    year = "2025"
}

@article{Banerjee:2022nbr,
    author = "Banerjee, Pulak and Coutinho, Antonio and Engel, Tim and Gurgone, Andrea and Signer, Adrian and Ulrich, Yannick",
    title = "{High-precision muon decay predictions for ALP searches}",
    eprint = "2211.01040",
    archivePrefix = "arXiv",
    primaryClass = "hep-ph",
    reportNumber = "FR-PHENO-2022-10, IFIC/22-32, IPPP/22/75, PSI-PR-22-32, ZU-TH 50/22",
    doi = "10.21468/SciPostPhys.15.1.021",
    journal = "SciPost Phys.",
    volume = "15",
    number = "1",
    pages = "021",
    year = "2023"
}

@article{Davidson:2022jai,
    author = "Davidson, Sacha and Echenard, Bertrand and Bernstein, Robert H. and Heeck, Julian and Hitlin, David G.",
    title = "{Charged Lepton Flavor Violation}",
    eprint = "2209.00142",
    archivePrefix = "arXiv",
    primaryClass = "hep-ex",
    month = "8",
    year = "2022"
}

@article{Bauer:2021mvw,
    author = "Bauer, Martin and Neubert, Matthias and Renner, Sophie and Schnubel, Marvin and Thamm, Andrea",
    title = "{Flavor probes of axion-like particles}",
    eprint = "2110.10698",
    archivePrefix = "arXiv",
    primaryClass = "hep-ph",
    reportNumber = "MITP/21-025, CERN-TH-2021-148, IPPP/21/37",
    doi = "10.1007/JHEP09(2022)056",
    journal = "JHEP",
    volume = "09",
    pages = "056",
    year = "2022"
}

@article{Belle:2025bpu,
    author = "Uno, K. and others",
    collaboration = "Belle",
    title = "{Search for lepton-flavor-violating tau decays to {\ensuremath{\ell}}{\ensuremath{\alpha}} at Belle}",
    eprint = "2503.22195",
    archivePrefix = "arXiv",
    primaryClass = "hep-ex",
    doi = "10.1007/JHEP08(2025)155",
    journal = "JHEP",
    volume = "08",
    pages = "155",
    year = "2025"
}

@article{TWIST:2014ymv,
    author = "Bayes, R. and others",
    collaboration = "TWIST",
    title = "{Search for two body muon decay signals}",
    eprint = "1409.0638",
    archivePrefix = "arXiv",
    primaryClass = "hep-ex",
    doi = "10.1103/PhysRevD.91.052020",
    journal = "Phys. Rev. D",
    volume = "91",
    number = "5",
    pages = "052020",
    year = "2015"
}

@article{Wang:2025xyh,
    author = "Wang, Zeren Simon and Zhang, Yu and Chen, Liangwen",
    title = "{Searching for long-lived particles from stopped pions and muons at the CiADS-BDE}",
    eprint = "2501.15460",
    archivePrefix = "arXiv",
    primaryClass = "hep-ph",
    month = "1",
    year = "2025"
}

@article{Knapen:2024fvh,
    author = "Knapen, Simon and Opferkuch, Toby and Redigolo, Diego and Tammaro, Michele",
    title = "{Displaced searches for Axion-Like Particles and Heavy Neutral Leptons at Mu3e}",
    eprint = "2410.13941",
    archivePrefix = "arXiv",
    primaryClass = "hep-ph",
    doi = "10.1007/JHEP06(2025)189",
    journal = "JHEP",
    volume = "06",
    pages = "189",
    year = "2025"
}

@article{Badziak:2024szg,
    author = "Badziak, Marcin and Harigaya, Keisuke and {\L}ukawski, Micha{\l} and Ziegler, Robert",
    title = "{Thermal production of astrophobic axions}",
    eprint = "2403.05621",
    archivePrefix = "arXiv",
    primaryClass = "hep-ph",
    doi = "10.1007/JHEP09(2024)136",
    journal = "JHEP",
    volume = "09",
    pages = "136",
    year = "2024"
}

@article{Knapen:2023zgi,
    author = "Knapen, Simon and Langhoff, Kevin and Opferkuch, Toby and Redigolo, Diego",
    title = "{A robust search for lepton flavour violating axions at Mu3e}",
    eprint = "2311.17915",
    archivePrefix = "arXiv",
    primaryClass = "hep-ph",
    doi = "10.1007/JHEP07(2025)243",
    journal = "JHEP",
    volume = "07",
    pages = "243",
    year = "2025"
}

@article{Hill:2023dym,
    author = "Hill, Richard J. and Plestid, Ryan and Zupan, Jure",
    title = "{Searching for new physics at $\mu \rightarrow e$ facilities with $\mu^+$ and $\pi^+$ decays at rest}",
    eprint = "2310.00043",
    archivePrefix = "arXiv",
    primaryClass = "hep-ph",
    reportNumber = "FERMILAB-PUB-23-287-T, CALT-TH/2023-017",
    doi = "10.1103/PhysRevD.109.035025",
    journal = "Phys. Rev. D",
    volume = "109",
    number = "3",
    pages = "035025",
    year = "2024"
}

@article{Belle-II:2022heu,
    author = "Adachi, I. and others",
    collaboration = "Belle-II",
    title = "{Search for Lepton-Flavor-Violating {\ensuremath{\tau}} Decays to a Lepton and an Invisible Boson at Belle II}",
    eprint = "2212.03634",
    archivePrefix = "arXiv",
    primaryClass = "hep-ex",
    reportNumber = "Belle II Preprint 2022-007, KEK Preprint 2022-39",
    doi = "10.1103/PhysRevLett.130.181803",
    journal = "Phys. Rev. Lett.",
    volume = "130",
    number = "18",
    pages = "181803",
    year = "2023"
}

@article{Han:2022iig,
    author = "Han, Chengcheng and L{\'o}pez-Ib{\'a}{\~n}ez, M. L. and Melis, Aurora and Vives, {\'O}scar and Yang, Jin Min",
    title = "{Anomaly-free ALP from non-Abelian flavor symmetry}",
    eprint = "2203.16376",
    archivePrefix = "arXiv",
    primaryClass = "hep-ph",
    reportNumber = "IFIC/22-12, FTUV-22-0331",
    doi = "10.1007/JHEP08(2022)306",
    journal = "JHEP",
    volume = "08",
    pages = "306",
    year = "2022"
}

@article{Jho:2022snj,
    author = "Jho, Yongsoo and Knapen, Simon and Redigolo, Diego",
    title = "{Lepton-flavor violating axions at MEG II}",
    eprint = "2203.11222",
    archivePrefix = "arXiv",
    primaryClass = "hep-ph",
    reportNumber = "CERN-TH-2022-044",
    doi = "10.1007/JHEP10(2022)029",
    journal = "JHEP",
    volume = "10",
    pages = "029",
    year = "2022"
}

@article{Guadagnoli:2021fcj,
    author = "Guadagnoli, Diego and Park, Chan Beom and Tenchini, Francesco",
    title = "{$\tau \rightarrow \ell$+invisible through invisible-savvy collider variables}",
    eprint = "2106.16236",
    archivePrefix = "arXiv",
    primaryClass = "hep-ph",
    reportNumber = "CERN-TH-2021-101, LAPTH-023/21, CTPU-PTC-21-28",
    doi = "10.1016/j.physletb.2021.136701",
    journal = "Phys. Lett. B",
    volume = "822",
    pages = "136701",
    year = "2021"
}

@article{Bryman:2021ilc,
    author = "Bryman, Douglas A. and Ito, Shintaro and Shrock, Robert",
    title = "{Upper limits on branching ratios of the lepton-flavor-violating decays $\tau \to \ell\gamma\gamma$ and $\tau \to \ell X$}",
    eprint = "2106.02451",
    archivePrefix = "arXiv",
    primaryClass = "hep-ph",
    doi = "10.1103/PhysRevD.104.075032",
    journal = "Phys. Rev. D",
    volume = "104",
    number = "7",
    pages = "075032",
    year = "2021"
}

@article{Ma:2021jkp,
    author = "Ma, Kai",
    title = "{Polarization and Correlation Effects in Lepton Flavor Violated Decays Induced by Axion-Like Particle}",
    eprint = "2104.11162",
    archivePrefix = "arXiv",
    primaryClass = "hep-ph",
    month = "4",
    year = "2021"
}

@article{Escribano:2020wua,
    author = "Escribano, Pablo and Vicente, Avelino",
    title = "{Ultralight scalars in leptonic observables}",
    eprint = "2008.01099",
    archivePrefix = "arXiv",
    primaryClass = "hep-ph",
    reportNumber = "IFIC/20-40",
    doi = "10.1007/JHEP03(2021)240",
    journal = "JHEP",
    volume = "03",
    pages = "240",
    year = "2021"
}

@article{Han:2020dwo,
    author = "Han, C. and L{\'o}pez-Ib{\'a}{\~n}ez, M. L. and Melis, A. and Vives, O. and Yang, J. M.",
    title = "{Anomaly-free leptophilic axionlike particle and its flavor violating tests}",
    eprint = "2007.08834",
    archivePrefix = "arXiv",
    primaryClass = "hep-ph",
    reportNumber = "FTUV-20-0717, IFIC/20-36",
    doi = "10.1103/PhysRevD.103.035028",
    journal = "Phys. Rev. D",
    volume = "103",
    number = "3",
    pages = "035028",
    year = "2021"
}

@article{DeLaCruz-Burelo:2020ozf,
    author = "De La Cruz-Burelo, E. and Hernandez-Villanueva, M. and De Yta-Hernandez, A.",
    title = "{New method for beyond the Standard Model invisible particle searches in tau lepton decays}",
    eprint = "2007.08239",
    archivePrefix = "arXiv",
    primaryClass = "hep-ph",
    doi = "10.1103/PhysRevD.102.115001",
    journal = "Phys. Rev. D",
    volume = "102",
    number = "11",
    pages = "115001",
    year = "2020"
}

@article{Jiang:2024cqj,
    author = "Jiang, Xu-Hui and Lu, Chih-Ting",
    title = "{Leptophilic axionlike particles at forward detectors}",
    eprint = "2412.19195",
    archivePrefix = "arXiv",
    primaryClass = "hep-ph",
    doi = "10.1103/PhysRevD.111.035035",
    journal = "Phys. Rev. D",
    volume = "111",
    number = "3",
    pages = "035035",
    year = "2025"
}

@article{Liu:2023bby,
    author = "Liu, Jia and Luo, Yan and Song, Muyuan",
    title = "{Investigation of the concurrent effects of ALP-photon and ALP-electron couplings in Collider and Beam Dump Searches}",
    eprint = "2304.05435",
    archivePrefix = "arXiv",
    primaryClass = "hep-ph",
    doi = "10.1007/JHEP09(2023)104",
    journal = "JHEP",
    volume = "09",
    pages = "104",
    year = "2023"
}

@article{Guerrera:2022ykl,
    author = "Guerrera, Alfredo Walter Mario and Rigolin, Stefano",
    title = "{ALP Production in Weak Mesonic Decays}",
    eprint = "2211.08343",
    archivePrefix = "arXiv",
    primaryClass = "hep-ph",
    doi = "10.1002/prop.202200192",
    journal = "Fortsch. Phys.",
    volume = "71",
    number = "2-3",
    pages = "2200192",
    year = "2023"
}

@article{Schafer:2022shi,
    author = {Sch{\"a}fer, Ruth and Tillinger, Finn and Westhoff, Susanne},
    title = "{Near or far detectors? A case study for long-lived particle searches at electron-positron colliders}",
    eprint = "2202.11714",
    archivePrefix = "arXiv",
    primaryClass = "hep-ph",
    reportNumber = "P3H-22-017",
    doi = "10.1103/PhysRevD.107.076022",
    journal = "Phys. Rev. D",
    volume = "107",
    number = "7",
    pages = "076022",
    year = "2023"
}

@article{Dobrich:2015jyk,
    author = {D{\"o}brich, Babette and Jaeckel, Joerg and Kahlhoefer, Felix and Ringwald, Andreas and Schmidt-Hoberg, Kai},
    title = "{ALPtraum: ALP production in proton beam dump experiments}",
    eprint = "1512.03069",
    archivePrefix = "arXiv",
    primaryClass = "hep-ph",
    reportNumber = "CERN-PH-TH-2015-293, DESY-15-237",
    doi = "10.1007/JHEP02(2016)018",
    journal = "JHEP",
    volume = "02",
    pages = "018",
    year = "2016"
}

@article{Dobrich:2019dxc,
    author = {D{\"o}brich, Babette and Jaeckel, Joerg and Spadaro, Tommaso},
    title = "{Light in the beam dump - ALP production from decay photons in proton beam-dumps}",
    eprint = "1904.02091",
    archivePrefix = "arXiv",
    primaryClass = "hep-ph",
    doi = "10.1007/JHEP05(2019)213",
    journal = "JHEP",
    volume = "05",
    pages = "213",
    year = "2019",
    note = "[Erratum: JHEP 10, 046 (2020)]"
}

@article{Blinov:2021say,
    author = "Blinov, Nikita and Kowalczyk, Elizabeth and Wynne, Margaret",
    title = "{Axion-like particle searches at DarkQuest}",
    eprint = "2112.09814",
    archivePrefix = "arXiv",
    primaryClass = "hep-ph",
    reportNumber = "FERMILAB-PUB-21-749-V",
    doi = "10.1007/JHEP02(2022)036",
    journal = "JHEP",
    volume = "02",
    pages = "036",
    year = "2022"
}

@article{Wang:2022ock,
    author = "Wang, Han and Yue, Chong-Xing and Guo, Yu-Chen and Cheng, Xue-Jia and Li, Xin-Yang",
    title = "{Prospects for searching for axion-like particles at the CEPC}",
    doi = "10.1088/1361-6471/ac8f61",
    journal = "J. Phys. G",
    volume = "49",
    number = "11",
    pages = "115002",
    year = "2022"
}

@article{Ema:2025bww,
    author = "Ema, Yohei and Fox, Patrick J. and Hostert, Matheus and Menzo, Tony and Pospelov, Maxim and Ray, Anupam and Zupan, Jure",
    title = "{Long-lived axionlike particles from tau decays}",
    eprint = "2507.15271",
    archivePrefix = "arXiv",
    primaryClass = "hep-ph",
    reportNumber = "CERN-TH-2025-123, FERMILAB-PUB-25-0408-T, N3AS-25-011",
    doi = "10.1103/51gx-m32t",
    journal = "Phys. Rev. D",
    volume = "112",
    number = "11",
    pages = "115028",
    year = "2025"
}

@article{Fiorillo:2025sln,
    author = "Fiorillo, Damiano F. G. and Pitik, Tetyana and Vitagliano, Edoardo",
    title = "{Supernova production of axionlike particles coupling to electrons, reloaded}",
    eprint = "2503.15630",
    archivePrefix = "arXiv",
    primaryClass = "hep-ph",
    doi = "10.1103/y1r2-gtb5",
    journal = "Phys. Rev. D",
    volume = "112",
    number = "8",
    pages = "083008",
    year = "2025",
    note = "[Erratum: Phys.Rev.D 113, 089902 (2026)]"
}

@article{ARGUS:1995bjh,
    author = "Albrecht, H. and others",
    collaboration = "ARGUS",
    title = "{A Search for lepton flavor violating decays $\tau \rightarrow e \alpha$, $\tau \rightarrow \mu \alpha$}",
    reportNumber = "DESY-95-071",
    doi = "10.1007/BF01579801",
    journal = "Z. Phys. C",
    volume = "68",
    pages = "25--28",
    year = "1995"
}

@article{Hoof:2022xbe,
    author = "Hoof, Sebastian and Schulz, Lena",
    title = "{Updated constraints on axion-like particles from temporal information in supernova SN1987A gamma-ray data}",
    eprint = "2212.09764",
    archivePrefix = "arXiv",
    primaryClass = "hep-ph",
    reportNumber = "TTP22-072",
    doi = "10.1088/1475-7516/2023/03/054",
    journal = "JCAP",
    volume = "03",
    pages = "054",
    year = "2023"
}

@article{Muller:2023vjm,
    author = {M{\"u}ller, Eike and Calore, Francesca and Carenza, Pierluca and Eckner, Christopher and Marsh, M. C. David},
    title = "{Investigating the gamma-ray burst from decaying MeV-scale axion-like particles produced in supernova explosions}",
    eprint = "2304.01060",
    archivePrefix = "arXiv",
    primaryClass = "astro-ph.HE",
    doi = "10.1088/1475-7516/2023/07/056",
    journal = "JCAP",
    volume = "07",
    pages = "056",
    year = "2023"
}

@article{Dev:2023hax,
    author = "Dev, P. S. Bhupal and Fortin, Jean-Fran{\c{c}}ois and Harris, Steven P. and Sinha, Kuver and Zhang, Yongchao",
    title = "{First Constraints on the Photon Coupling of Axionlike Particles from Multimessenger Studies of the Neutron Star Merger GW170817}",
    eprint = "2305.01002",
    archivePrefix = "arXiv",
    primaryClass = "hep-ph",
    reportNumber = "INT-PUB-23-014",
    doi = "10.1103/PhysRevLett.132.101003",
    journal = "Phys. Rev. Lett.",
    volume = "132",
    number = "10",
    pages = "101003",
    year = "2024"
}

@article{Croon:2020lrf,
    author = "Croon, Djuna and Elor, Gilly and Leane, Rebecca K. and McDermott, Samuel D.",
    title = "{Supernova Muons: New Constraints on $Z$' Bosons, Axions and ALPs}",
    eprint = "2006.13942",
    archivePrefix = "arXiv",
    primaryClass = "hep-ph",
    reportNumber = "MIT-CTP/5214, FERMILAB-PUB-20-246-A-T",
    doi = "10.1007/JHEP01(2021)107",
    journal = "JHEP",
    volume = "01",
    pages = "107",
    year = "2021"
}

@article{Bolton:1988af,
    author = "Bolton, R. D. and others",
    title = "{Search for Rare Muon Decays with the Crystal Box Detector}",
    reportNumber = "LA-UR-88-392",
    doi = "10.1103/PhysRevD.38.2077",
    journal = "Phys. Rev. D",
    volume = "38",
    pages = "2077",
    year = "1988"
}

@article{PIENU:2020loi,
    author = "Aguilar-Arevalo, A. and others",
    collaboration = "PIENU",
    title = "{Improved search for two body muon decay ${\mu}^+{\rightarrow}e^+X_H$}",
    eprint = "2002.09170",
    archivePrefix = "arXiv",
    primaryClass = "hep-ex",
    doi = "10.1103/PhysRevD.101.052014",
    journal = "Phys. Rev. D",
    volume = "101",
    number = "5",
    pages = "052014",
    year = "2020"
}

@inproceedings{PIONEER:2022alm,
    author = "Altmannshofer, W. and others",
    collaboration = "PIONEER",
    title = "{Testing Lepton Flavor Universality and CKM Unitarity with Rare Pion Decays in the PIONEER experiment}",
    booktitle = "{Snowmass 2021}",
    eprint = "2203.05505",
    archivePrefix = "arXiv",
    primaryClass = "hep-ex",
    reportNumber = "FERMILAB-CONF-22-328-PPD",
    month = "3",
    year = "2022"
}

@article{Dev:2016dja,
    author = "Dev, P. S. Bhupal and Mohapatra, Rabindra N. and Zhang, Yongchao",
    title = "{Probing the Higgs Sector of the Minimal Left-Right Symmetric Model at Future Hadron Colliders}",
    eprint = "1602.05947",
    archivePrefix = "arXiv",
    primaryClass = "hep-ph",
    reportNumber = "UMD-PP-016-002, ULB-TH-16-03",
    doi = "10.1007/JHEP05(2016)174",
    journal = "JHEP",
    volume = "05",
    pages = "174",
    year = "2016"
}

@article{BhupalDev:2016nfr,
    author = "Bhupal Dev, P. S. and Mohapatra, Rabindra N. and Zhang, Yongchao",
    title = "{Displaced photon signal from a possible light scalar in minimal left-right seesaw model}",
    eprint = "1612.09587",
    archivePrefix = "arXiv",
    primaryClass = "hep-ph",
    reportNumber = "ULB-TH-17-11",
    doi = "10.1103/PhysRevD.95.115001",
    journal = "Phys. Rev. D",
    volume = "95",
    number = "11",
    pages = "115001",
    year = "2017"
}

@article{Jaeckel:2015jla,
    author = "Jaeckel, Joerg and Spannowsky, Michael",
    title = "{Probing MeV to 90 GeV axion-like particles with LEP and LHC}",
    eprint = "1509.00476",
    archivePrefix = "arXiv",
    primaryClass = "hep-ph",
    doi = "10.1016/j.physletb.2015.12.037",
    journal = "Phys. Lett. B",
    volume = "753",
    pages = "482--487",
    year = "2016"
}

@article{Bauer:2018uxu,
    author = "Bauer, Martin and Heiles, Mathias and Neubert, Matthias and Thamm, Andrea",
    title = "{Axion-Like Particles at Future Colliders}",
    eprint = "1808.10323",
    archivePrefix = "arXiv",
    primaryClass = "hep-ph",
    reportNumber = "CERN-TH-2018-199, MITP/18-075",
    doi = "10.1140/epjc/s10052-019-6587-9",
    journal = "Eur. Phys. J. C",
    volume = "79",
    number = "1",
    pages = "74",
    year = "2019"
}

@article{Bauer:2017ris,
    author = "Bauer, Martin and Neubert, Matthias and Thamm, Andrea",
    title = "{Collider Probes of Axion-Like Particles}",
    eprint = "1708.00443",
    archivePrefix = "arXiv",
    primaryClass = "hep-ph",
    reportNumber = "MITP-17-047",
    doi = "10.1007/JHEP12(2017)044",
    journal = "JHEP",
    volume = "12",
    pages = "044",
    year = "2017"
}

@article{Brivio:2017ije,
    author = "Brivio, I. and Gavela, M. B. and Merlo, L. and Mimasu, K. and No, J. M. and del Rey, R. and Sanz, V.",
    title = "{ALPs Effective Field Theory and Collider Signatures}",
    eprint = "1701.05379",
    archivePrefix = "arXiv",
    primaryClass = "hep-ph",
    reportNumber = "IFT-UAM-CSIC-16-141, KCL-PH-TH-2016-72, FTUAM-16-49, CP3-17-04",
    doi = "10.1140/epjc/s10052-017-5111-3",
    journal = "Eur. Phys. J. C",
    volume = "77",
    number = "8",
    pages = "572",
    year = "2017"
}

@article{Alekhin:2015byh,
    author = "Alekhin, Sergey and others",
    title = "{A facility to Search for Hidden Particles at the CERN SPS: the SHiP physics case}",
    eprint = "1504.04855",
    archivePrefix = "arXiv",
    primaryClass = "hep-ph",
    reportNumber = "CERN-SPSC-2015-017, SPSC-P-350-ADD-1",
    doi = "10.1088/0034-4885/79/12/124201",
    journal = "Rept. Prog. Phys.",
    volume = "79",
    number = "12",
    pages = "124201",
    year = "2016"
}

@article{Feng:2018pew,
    author = "Feng, Jonathan L. and Galon, Iftah and Kling, Felix and Trojanowski, Sebastian",
    title = "{Axionlike particles at FASER: The LHC as a photon beam dump}",
    eprint = "1806.02348",
    archivePrefix = "arXiv",
    primaryClass = "hep-ph",
    reportNumber = "UCI-TR-2018-02",
    doi = "10.1103/PhysRevD.98.055021",
    journal = "Phys. Rev. D",
    volume = "98",
    number = "5",
    pages = "055021",
    year = "2018"
}

@article{Mimasu:2014nea,
    author = "Mimasu, Ken and Sanz, Ver{\'o}nica",
    title = "{ALPs at Colliders}",
    eprint = "1409.4792",
    archivePrefix = "arXiv",
    primaryClass = "hep-ph",
    doi = "10.1007/JHEP06(2015)173",
    journal = "JHEP",
    volume = "06",
    pages = "173",
    year = "2015"
}

@article{NA64:2020qwq,
    author = "Banerjee, D. and others",
    collaboration = "NA64",
    title = "{Search for Axionlike and Scalar Particles with the NA64 Experiment}",
    eprint = "2005.02710",
    archivePrefix = "arXiv",
    primaryClass = "hep-ex",
    reportNumber = "CERN-EP-2020-068",
    doi = "10.1103/PhysRevLett.125.081801",
    journal = "Phys. Rev. Lett.",
    volume = "125",
    number = "8",
    pages = "081801",
    year = "2020"
}

@article{Perrevoort:2018ttp,
    author = "Perrevoort, Ann-Kathrin",
    collaboration = "Mu3e",
    title = "{The Rare and Forbidden: Testing Physics Beyond the Standard Model with Mu3e}",
    eprint = "1812.00741",
    archivePrefix = "arXiv",
    primaryClass = "hep-ex",
    doi = "10.21468/SciPostPhysProc.1.052",
    journal = "SciPost Phys. Proc.",
    volume = "1",
    pages = "052",
    year = "2019"
}

@article{Dusaev:2020gxi,
    author = "Dusaev, R. R. and Kirpichnikov, D. V. and Kirsanov, M. M.",
    title = "{Photoproduction of axionlike particles in the NA64 experiment}",
    eprint = "2004.04469",
    archivePrefix = "arXiv",
    primaryClass = "hep-ph",
    doi = "10.1103/PhysRevD.102.055018",
    journal = "Phys. Rev. D",
    volume = "102",
    number = "5",
    pages = "055018",
    year = "2020"
}

@article{Bai:2021gbm,
    author = "Bai, Zhaoyu and others",
    title = "{New physics searches with an optical dump at LUXE}",
    eprint = "2107.13554",
    archivePrefix = "arXiv",
    primaryClass = "hep-ph",
    reportNumber = "DESY 21-111",
    doi = "10.1103/PhysRevD.106.115034",
    journal = "Phys. Rev. D",
    volume = "106",
    number = "11",
    pages = "115034",
    year = "2022"
}

@article{Kriewald:2024cgr,
    author = "Kriewald, Jonathan and Nemev{\v{s}}ek, Miha and Nesti, Fabrizio",
    title = "{Enabling precise predictions for left-right symmetry at colliders}",
    eprint = "2403.07756",
    archivePrefix = "arXiv",
    primaryClass = "hep-ph",
    doi = "10.1140/epjc/s10052-024-13614-8",
    journal = "Eur. Phys. J. C",
    volume = "84",
    number = "12",
    pages = "1306",
    year = "2024"
}

@article{Cirigliano:2004mv,
    author = "Cirigliano, V. and Kurylov, A. and Ramsey-Musolf, M. J. and Vogel, P.",
    title = "{Lepton flavor violation without supersymmetry}",
    eprint = "hep-ph/0404233",
    archivePrefix = "arXiv",
    reportNumber = "MAP-296",
    doi = "10.1103/PhysRevD.70.075007",
    journal = "Phys. Rev. D",
    volume = "70",
    pages = "075007",
    year = "2004"
}

@article{Das:2012ii,
    author = "Das, S. P. and Deppisch, F. F. and Kittel, O. and Valle, J. W. F.",
    title = "{Heavy Neutrinos and Lepton Flavour Violation in Left-Right Symmetric Models at the LHC}",
    eprint = "1206.0256",
    archivePrefix = "arXiv",
    primaryClass = "hep-ph",
    reportNumber = "IFIC-12-17",
    doi = "10.1103/PhysRevD.86.055006",
    journal = "Phys. Rev. D",
    volume = "86",
    pages = "055006",
    year = "2012"
}

@article{Aguilar-Saavedra:2012dga,
    author = "Aguilar-Saavedra, J. A. and Deppisch, F. and Kittel, O. and Valle, J. W. F.",
    title = "{Flavour in heavy neutrino searches at the LHC}",
    eprint = "1203.5998",
    archivePrefix = "arXiv",
    primaryClass = "hep-ph",
    reportNumber = "IFIC-12-18",
    doi = "10.1103/PhysRevD.85.091301",
    journal = "Phys. Rev. D",
    volume = "85",
    pages = "091301",
    year = "2012"
}

@article{Belle:2000cnh,
    author = "Abashian, A. and others",
    collaboration = "Belle",
    title = "{The Belle Detector}",
    reportNumber = "KEK-PROGRESS-REPORT-2000-4",
    doi = "10.1016/S0168-9002(01)02013-7",
    journal = "Nucl. Instrum. Meth. A",
    volume = "479",
    pages = "117--232",
    year = "2002"
}

@article{Greljo:2025ljr,
    author = "Greljo, Admir and Palavri{\'c}, Ajdin and Tunja, Mirsad and Zupan, Jure",
    title = "{Expanding the landscape of exotic muon decays}",
    eprint = "2510.08674",
    archivePrefix = "arXiv",
    primaryClass = "hep-ph",
    doi = "10.1103/cb52-r75c",
    journal = "Phys. Rev. D",
    volume = "113",
    number = "7",
    pages = "075022",
    year = "2026"
}

@article{Jaeckel:2017tud,
    author = "Jaeckel, J. and Malta, P. C. and Redondo, J.",
    title = "{Decay photons from the axionlike particles burst of type II supernovae}",
    eprint = "1702.02964",
    archivePrefix = "arXiv",
    primaryClass = "hep-ph",
    doi = "10.1103/PhysRevD.98.055032",
    journal = "Phys. Rev. D",
    volume = "98",
    number = "5",
    pages = "055032",
    year = "2018"
}

@article{Ferreira:2025qui,
    author = "Ferreira, Ricardo Z. and Marsh, M. C. David and Ravensburg, Eike",
    title = "{ALP couplings to muons and electrons: a comprehensive analysis of supernova bounds}",
    eprint = "2510.14469",
    archivePrefix = "arXiv",
    primaryClass = "hep-ph",
    month = "10",
    year = "2025"
}

@article{Jodidio:1986mz,
    author = "Jodidio, A. and others",
    title = "{Search for Right-Handed Currents in Muon Decay}",
    reportNumber = "LBL-21616",
    doi = "10.1103/PhysRevD.34.1967",
    journal = "Phys. Rev. D",
    volume = "34",
    pages = "1967",
    year = "1986",
    note = "[Erratum: Phys.Rev.D 37, 237 (1988)]"
}

@article{Blanke:2011ry,
    author = "Blanke, Monika and Buras, Andrzej J. and Gemmler, Katrin and Heidsieck, Tillmann",
    title = "{Delta F = 2 observables and $B \to X_q \gamma$ decays in the Left-Right Model: Higgs particles striking back}",
    eprint = "1111.5014",
    archivePrefix = "arXiv",
    primaryClass = "hep-ph",
    reportNumber = "TUM-HEP-820-11, FLAVOUR(267104)-ERC-5",
    doi = "10.1007/JHEP03(2012)024",
    journal = "JHEP",
    volume = "03",
    pages = "024",
    year = "2012"
}

@article{Zhang:2007da,
    author = "Zhang, Yue and An, Haipeng and Ji, Xiangdong and Mohapatra, Rabindra N.",
    title = "{General CP Violation in Minimal Left-Right Symmetric Model and Constraints on the Right-Handed Scale}",
    eprint = "0712.4218",
    archivePrefix = "arXiv",
    primaryClass = "hep-ph",
    doi = "10.1016/j.nuclphysb.2008.05.019",
    journal = "Nucl. Phys. B",
    volume = "802",
    pages = "247--279",
    year = "2008"
}

@article{Chugai:1999en,
    author = "Chugai, Nikolai N. and Utrobin, Victor P.",
    title = "{The nature of sn 1997d: low mass progenitor and weak explosion}",
    eprint = "astro-ph/9906190",
    archivePrefix = "arXiv",
    journal = "Astron. Astrophys.",
    volume = "354",
    pages = "557",
    year = "2000"
}

@article{Pastorello:2003tc,
    author = "Pastorello, A. and others",
    title = "{Low luminosity type II supernovae: spectroscopic and photometric evolution}",
    eprint = "astro-ph/0309264",
    archivePrefix = "arXiv",
    doi = "10.1111/j.1365-2966.2004.07173.x",
    journal = "Mon. Not. Roy. Astron. Soc.",
    volume = "347",
    pages = "74",
    year = "2004"
}

@article{Burrows:2020qrp,
    author = "Burrows, Adam and Vartanyan, David",
    title = "{Core-Collapse Supernova Explosion Theory}",
    eprint = "2009.14157",
    archivePrefix = "arXiv",
    primaryClass = "astro-ph.SR",
    doi = "10.1038/s41586-020-03059-w",
    journal = "Nature",
    volume = "589",
    number = "7840",
    pages = "29--39",
    year = "2021"
}

@article{Kitaura:2005bt,
    author = "Kitaura, F. S. and Janka, Hans-Thomas and Hillebrandt, W.",
    title = "{Explosions of O-Ne-Mg cores, the Crab supernova, and subluminous type II-P supernovae}",
    eprint = "astro-ph/0512065",
    archivePrefix = "arXiv",
    doi = "10.1051/0004-6361:20054703",
    journal = "Astron. Astrophys.",
    volume = "450",
    pages = "345--350",
    year = "2006"
}

@article{Melson:2015tia,
    author = "Melson, Tobias and Janka, Hans-Thomas and Marek, Andreas",
    title = "{Neutrino-driven supernova of a low-mass iron-core progenitor boosted by three-dimensional turbulent convection}",
    eprint = "1501.01961",
    archivePrefix = "arXiv",
    primaryClass = "astro-ph.SR",
    doi = "10.1088/2041-8205/801/2/L24",
    journal = "Astrophys. J. Lett.",
    volume = "801",
    number = "2",
    pages = "L24",
    year = "2015"
}

@article{Radice:2017ykv,
    author = "Radice, David and Burrows, Adam and Vartanyan, David and Skinner, M. Aaron and Dolence, Joshua C.",
    title = "{Electron-Capture and Low-Mass Iron-Core-Collapse Supernovae: New Neutrino-Radiation-Hydrodynamics Simulations}",
    eprint = "1702.03927",
    archivePrefix = "arXiv",
    primaryClass = "astro-ph.HE",
    reportNumber = "LA-UR-17-20973",
    doi = "10.3847/1538-4357/aa92c5",
    journal = "Astrophys. J.",
    volume = "850",
    number = "1",
    pages = "43",
    year = "2017"
}

@article{Muller:2018utr,
    author = {M{\"u}ller, B. and Tauris, T. M. and Heger, A. and Banerjee, P. and Qian, Y. -Z. and Powell, J. and Chan, C. and Gay, D. W. and Langer, N.},
    title = "{Three-Dimensional Simulations of Neutrino-Driven Core-Collapse Supernovae from Low-Mass Single and Binary Star Progenitors}",
    eprint = "1811.05483",
    archivePrefix = "arXiv",
    primaryClass = "astro-ph.SR",
    doi = "10.1093/mnras/stz216",
    journal = "Mon. Not. Roy. Astron. Soc.",
    volume = "484",
    number = "3",
    pages = "3307--3324",
    year = "2019"
}

@article{Burrows:2019rtd,
    author = "Burrows, Adam and Radice, David and Vartanyan, David",
    title = "{Three-dimensional supernova explosion simulations of 9-, 10-, 11-, 12-, and 13-M{\ensuremath{\odot}} stars}",
    eprint = "1902.00547",
    archivePrefix = "arXiv",
    primaryClass = "astro-ph.SR",
    doi = "10.1093/mnras/stz543",
    journal = "Mon. Not. Roy. Astron. Soc.",
    volume = "485",
    number = "3",
    pages = "3153--3168",
    year = "2019"
}

@article{Stockinger:2020hse,
    author = "Stockinger, G. and others",
    title = "{Three-dimensional Models of Core-collapse Supernovae From Low-mass Progenitors With Implications for Crab}",
    eprint = "2005.02420",
    archivePrefix = "arXiv",
    primaryClass = "astro-ph.HE",
    doi = "10.1093/mnras/staa1691",
    journal = "Mon. Not. Roy. Astron. Soc.",
    volume = "496",
    number = "2",
    pages = "2039--2084",
    year = "2020"
}

@article{Carenza:2021pcm,
    author = "Carenza, Pierluca and Lucente, Giuseppe",
    title = "{Supernova bound on axionlike particles coupled with electrons}",
    eprint = "2107.12393",
    archivePrefix = "arXiv",
    primaryClass = "hep-ph",
    doi = "10.1103/PhysRevD.104.103007",
    journal = "Phys. Rev. D",
    volume = "104",
    number = "10",
    pages = "103007",
    year = "2021",
    note = "[Erratum: Phys.Rev.D 110, 049901 (2024)]"
}

@article{Ferreira:2022xlw,
    author = {Ferreira, Ricardo Z. and Marsh, M. C. David and M{\"u}ller, Eike},
    title = "{Strong supernovae bounds on ALPs from quantum loops}",
    eprint = "2205.07896",
    archivePrefix = "arXiv",
    primaryClass = "hep-ph",
    doi = "10.1088/1475-7516/2022/11/057",
    journal = "JCAP",
    volume = "11",
    pages = "057",
    year = "2022"
}

@article{Huang:2025xvo,
    author = "Huang, Zi-Miao and Li, Changqian and Liu, Zuowei",
    title = "{Refined Low-Energy Supernova Constraints on Lepton Flavor Violating Axions}",
    eprint = "2510.22523",
    archivePrefix = "arXiv",
    primaryClass = "hep-ph",
    month = "10",
    year = "2025"
}

@article{Maiezza:2016ybz,
    author = "Maiezza, Alessio and Senjanovi{\'c}, Goran and Vasquez, Juan Carlos",
    title = "{Higgs sector of the minimal left-right symmetric theory}",
    eprint = "1612.09146",
    archivePrefix = "arXiv",
    primaryClass = "hep-ph",
    doi = "10.1103/PhysRevD.95.095004",
    journal = "Phys. Rev. D",
    volume = "95",
    number = "9",
    pages = "095004",
    year = "2017"
}

@article{EscuderoAbenza:2025tsi,
    author = "Escudero Abenza, Miguel and Garcia-Perez, Clara and Ovchynnikov, Maksym",
    title = "{Nucleosynthesis and CMB bounds on photophilic ALPs: a fresh look}",
    eprint = "2511.00157",
    archivePrefix = "arXiv",
    primaryClass = "hep-ph",
    reportNumber = "CERN-TH-2025-222",
    doi = "10.1140/epjc/s10052-026-15544-z",
    journal = "Eur. Phys. J. C",
    volume = "86",
    number = "5",
    pages = "500",
    year = "2026"
}

@article{Diamond:2023cto,
    author = "Diamond, Melissa and Fiorillo, Damiano F. G. and Marques-Tavares, Gustavo and Tamborra, Irene and Vitagliano, Edoardo",
    title = "{Multimessenger Constraints on Radiatively Decaying Axions from GW170817}",
    eprint = "2305.10327",
    archivePrefix = "arXiv",
    primaryClass = "hep-ph",
    doi = "10.1103/PhysRevLett.132.101004",
    journal = "Phys. Rev. Lett.",
    volume = "132",
    number = "10",
    pages = "101004",
    year = "2024"
}

@article{Diamond:2023scc,
    author = "Diamond, Melissa and Fiorillo, Damiano F. G. and Marques-Tavares, Gustavo and Vitagliano, Edoardo",
    title = "{Axion-sourced fireballs from supernovae}",
    eprint = "2303.11395",
    archivePrefix = "arXiv",
    primaryClass = "hep-ph",
    doi = "10.1103/PhysRevD.107.103029",
    journal = "Phys. Rev. D",
    volume = "107",
    number = "10",
    pages = "103029",
    year = "2023",
    note = "[Erratum: Phys.Rev.D 108, 049902 (2023)]"
}

@article{Capozzi:2023ffu,
    author = "Capozzi, Francesco and Dutta, Bhaskar and Gurung, Gajendra and Jang, Wooyoung and Shoemaker, Ian M. and Thompson, Adrian and Yu, Jaehoon",
    title = "{New constraints on ALP couplings to electrons and photons from ArgoNeuT and the MiniBooNE beam dump}",
    eprint = "2307.03878",
    archivePrefix = "arXiv",
    primaryClass = "hep-ph",
    reportNumber = "MI-HET-808",
    doi = "10.1103/PhysRevD.108.075019",
    journal = "Phys. Rev. D",
    volume = "108",
    number = "7",
    pages = "075019",
    year = "2023"
}

@article{Caputo:2022mah,
    author = "Caputo, Andrea and Janka, Hans-Thomas and Raffelt, Georg and Vitagliano, Edoardo",
    title = "{Low-Energy Supernovae Severely Constrain Radiative Particle Decays}",
    eprint = "2201.09890",
    archivePrefix = "arXiv",
    primaryClass = "astro-ph.HE",
    doi = "10.1103/PhysRevLett.128.221103",
    journal = "Phys. Rev. Lett.",
    volume = "128",
    number = "22",
    pages = "221103",
    year = "2022"
}

@article{Calore:2021klc,
    author = "Calore, Francesca and Carenza, Pierluca and Giannotti, Maurizio and Jaeckel, Joerg and Lucente, Giuseppe and Mirizzi, Alessandro",
    title = "{Supernova bounds on axionlike particles coupled with nucleons and electrons}",
    eprint = "2107.02186",
    archivePrefix = "arXiv",
    primaryClass = "hep-ph",
    doi = "10.1103/PhysRevD.104.043016",
    journal = "Phys. Rev. D",
    volume = "104",
    number = "4",
    pages = "043016",
    year = "2021"
}

@article{Das:2017hmg,
    author = "Das, Arindam and Dev, P. S. Bhupal and Mohapatra, Rabindra N.",
    title = "{Same Sign versus Opposite Sign Dileptons as a Probe of Low Scale Seesaw Mechanisms}",
    eprint = "1709.06553",
    archivePrefix = "arXiv",
    primaryClass = "hep-ph",
    reportNumber = "UMD-PP-017-30",
    doi = "10.1103/PhysRevD.97.015018",
    journal = "Phys. Rev. D",
    volume = "97",
    number = "1",
    pages = "015018",
    year = "2018"
}

@article{Nemevsek:2016enw,
    author = "Nemev{\v{s}}ek, Miha and Nesti, Fabrizio and Vasquez, Juan Carlos",
    title = "{Majorana Higgses at colliders}",
    eprint = "1612.06840",
    archivePrefix = "arXiv",
    primaryClass = "hep-ph",
    doi = "10.1007/JHEP04(2017)114",
    journal = "JHEP",
    volume = "04",
    pages = "114",
    year = "2017"
}

@article{Bonilla:2016fqd,
    author = "Bonilla, Cesar and Krauss, Manuel E. and Opferkuch, Toby and Porod, Werner",
    title = "{Perspectives for Detecting Lepton Flavour Violation in Left-Right Symmetric Models}",
    eprint = "1611.07025",
    archivePrefix = "arXiv",
    primaryClass = "hep-ph",
    reportNumber = "BONN-TH-2016-08, IFIC-16-XX",
    doi = "10.1007/JHEP03(2017)027",
    journal = "JHEP",
    volume = "03",
    pages = "027",
    year = "2017"
}

@article{Alonso:2012ji,
    author = "Alonso, R. and Dhen, M. and Gavela, M. B. and Hambye, T.",
    title = "{Muon conversion to electron in nuclei in type-I seesaw models}",
    eprint = "1209.2679",
    archivePrefix = "arXiv",
    primaryClass = "hep-ph",
    reportNumber = "ULB-TH-12-12, FTUAM-12-100, IFT-UAM-CSIC-12-78",
    doi = "10.1007/JHEP01(2013)118",
    journal = "JHEP",
    volume = "01",
    pages = "118",
    year = "2013"
}

@article{Dinh:2012bp,
    author = "Dinh, D. N. and Ibarra, A. and Molinaro, E. and Petcov, S. T.",
    title = "{The $\mu - e$ Conversion in Nuclei, $\mu \to e \gamma, \mu \to 3e$ Decays and TeV Scale See-Saw Scenarios of Neutrino Mass Generation}",
    eprint = "1205.4671",
    archivePrefix = "arXiv",
    primaryClass = "hep-ph",
    reportNumber = "FLAVOUR(267104)-ERC-15, SISSA-10-2012-EP, TUM-HEP-837-12, CFTP-12-007",
    doi = "10.1007/JHEP08(2012)125",
    journal = "JHEP",
    volume = "08",
    pages = "125",
    year = "2012",
    note = "[Erratum: JHEP 09, 023 (2013)]"
}

@article{Tello:2010am,
    author = "Tello, Vladimir and Nemevsek, Miha and Nesti, Fabrizio and Senjanovic, Goran and Vissani, Francesco",
    title = "{Left-Right Symmetry: from LHC to Neutrinoless Double Beta Decay}",
    eprint = "1011.3522",
    archivePrefix = "arXiv",
    primaryClass = "hep-ph",
    doi = "10.1103/PhysRevLett.106.151801",
    journal = "Phys. Rev. Lett.",
    volume = "106",
    pages = "151801",
    year = "2011"
}

@article{Lindner:2016bgg,
    author = "Lindner, Manfred and Platscher, Moritz and Queiroz, Farinaldo S.",
    title = "{A Call for New Physics : The Muon Anomalous Magnetic Moment and Lepton Flavor Violation}",
    eprint = "1610.06587",
    archivePrefix = "arXiv",
    primaryClass = "hep-ph",
    doi = "10.1016/j.physrep.2017.12.001",
    journal = "Phys. Rept.",
    volume = "731",
    pages = "1--82",
    year = "2018"
}

@article{Akeroyd:2006bb,
    author = "Akeroyd, A. G. and Aoki, Mayumi and Okada, Yasuhiro",
    title = "{Lepton Flavour Violating tau Decays in the Left-Right Symmetric Model}",
    eprint = "hep-ph/0610344",
    archivePrefix = "arXiv",
    reportNumber = "KEK-TH-1106",
    doi = "10.1103/PhysRevD.76.013004",
    journal = "Phys. Rev. D",
    volume = "76",
    pages = "013004",
    year = "2007"
}

@article{Keung:1983uu,
    author = "Keung, Wai-Yee and Senjanovic, Goran",
    title = "{Majorana Neutrinos and the Production of the Right-handed Charged Gauge Boson}",
    reportNumber = "BNL-32872",
    doi = "10.1103/PhysRevLett.50.1427",
    journal = "Phys. Rev. Lett.",
    volume = "50",
    pages = "1427",
    year = "1983"
}

@article{Dev:2025fcv,
    author = "Dev, P. S. Bhupal and Heeck, Julian and Thapa, Anil",
    title = "{Decaying scalar dark matter in the minimal left-right symmetric model}",
    eprint = "2501.14669",
    archivePrefix = "arXiv",
    primaryClass = "hep-ph",
    doi = "10.1103/x6x1-76p8",
    journal = "Phys. Rev. D",
    volume = "112",
    number = "1",
    pages = "015004",
    year = "2025"
}

@article{ATLAS:2023cjo,
    author = "Aad, Georges and others",
    collaboration = "ATLAS",
    title = "{Search for heavy Majorana or Dirac neutrinos and right-handed W gauge bosons in final states with charged leptons and jets in pp collisions at $\sqrt{s}=13$~TeV with the ATLAS detector}",
    eprint = "2304.09553",
    archivePrefix = "arXiv",
    primaryClass = "hep-ex",
    reportNumber = "CERN-EP-2023-034",
    doi = "10.1140/epjc/s10052-023-12021-9",
    journal = "Eur. Phys. J. C",
    volume = "83",
    number = "12",
    pages = "1164",
    year = "2023"
}

@article{CMS:2021dzb,
    author = "Tumasyan, Armen and others",
    collaboration = "CMS",
    title = "{Search for a right-handed W boson and a heavy neutrino in proton-proton collisions at $ \sqrt{s} $ = 13 TeV}",
    eprint = "2112.03949",
    archivePrefix = "arXiv",
    primaryClass = "hep-ex",
    reportNumber = "CMS-EXO-20-002, CERN-EP-2021-228",
    doi = "10.1007/JHEP04(2022)047",
    journal = "JHEP",
    volume = "04",
    pages = "047",
    year = "2022"
}

@article{Chauhan:2018uuy,
    author = "Chauhan, Garv and Dev, P. S. Bhupal and Mohapatra, Rabindra N. and Zhang, Yongchao",
    title = "{Perturbativity constraints on $U(1)_{B-L}$ and left-right models and implications for heavy gauge boson searches}",
    eprint = "1811.08789",
    archivePrefix = "arXiv",
    primaryClass = "hep-ph",
    doi = "10.1007/JHEP01(2019)208",
    journal = "JHEP",
    volume = "01",
    pages = "208",
    year = "2019"
}

@article{Roitgrund:2020cge,
    author = "Roitgrund, Aviad and Eilam, Gad",
    title = "{Search for like-sign dileptons plus two jets signal in the framework of the manifest left-right symmetric model}",
    eprint = "1704.07772",
    archivePrefix = "arXiv",
    primaryClass = "hep-ph",
    doi = "10.1007/JHEP01(2021)031",
    journal = "JHEP",
    volume = "01",
    pages = "031",
    year = "2021",
    note = "[Erratum: JHEP 03, 029 (2021)]"
}

@article{ParticleDataGroup:2024cfk,
    author = "Navas, S. and others",
    collaboration = "Particle Data Group",
    title = "{Review of particle physics}",
    doi = "10.1103/PhysRevD.110.030001",
    journal = "Phys. Rev. D",
    volume = "110",
    number = "3",
    pages = "030001",
    year = "2024"
}

@article{Mohapatra:1980yp,
    author = "Mohapatra, Rabindra N. and Senjanovic, Goran",
    title = "{Neutrino Masses and Mixings in Gauge Models with Spontaneous Parity Violation}",
    reportNumber = "FERMILAB-PUB-80-061-THY, FERMILAB-PUB-80-061-T",
    doi = "10.1103/PhysRevD.23.165",
    journal = "Phys. Rev. D",
    volume = "23",
    pages = "165",
    year = "1981"
}

@article{Magg:1980ut,
    author = "Magg, M. and Wetterich, C.",
    title = "{Neutrino Mass Problem and Gauge Hierarchy}",
    reportNumber = "CERN-TH-2829",
    doi = "10.1016/0370-2693(80)90825-4",
    journal = "Phys. Lett. B",
    volume = "94",
    pages = "61--64",
    year = "1980"
}

@article{Schechter:1980gr,
    author = "Schechter, J. and Valle, J. W. F.",
    title = "{Neutrino Masses in $SU(2) \times U(1)$ Theories}",
    reportNumber = "SU-4217-167, COO-3533-167",
    doi = "10.1103/PhysRevD.22.2227",
    journal = "Phys. Rev. D",
    volume = "22",
    pages = "2227",
    year = "1980"
}

@article{Cheng:1980qt,
    author = "Cheng, T. P. and Li, Ling-Fong",
    title = "{Neutrino Masses, Mixings and Oscillations in $SU(2) \times U(1)$ Models of Electroweak Interactions}",
    reportNumber = "PRINT-80-0511 (CARNEGIE-MELLON), COO-3066-152",
    doi = "10.1103/PhysRevD.22.2860",
    journal = "Phys. Rev. D",
    volume = "22",
    pages = "2860",
    year = "1980"
}

@article{Lazarides:1980nt,
    author = "Lazarides, George and Shafi, Q. and Wetterich, C.",
    title = "{Proton Lifetime and Fermion Masses in an SO(10) Model}",
    reportNumber = "FREIBURG-THEP-80-2",
    doi = "10.1016/0550-3213(81)90354-0",
    journal = "Nucl. Phys. B",
    volume = "181",
    pages = "287--300",
    year = "1981"
}

@article{Badziak:2025mkt,
    author = "Badziak, Marcin and Gomu{\l}ka, Adam and Laletin, Maxim and Szafra{\'n}ski, Krzysztof",
    title = "{Improved cosmological constraints on axion-lepton interactions}",
    eprint = "2511.14864",
    archivePrefix = "arXiv",
    primaryClass = "hep-ph",
    month = "11",
    year = "2025"
}

@article{BhupalDev:2018vpr,
    author = "Bhupal Dev, P. S. and Mohapatra, Rabindra N. and Zhang, Yongchao",
    title = "{Probing TeV scale origin of neutrino mass at future lepton colliders via neutral and doubly-charged scalars}",
    eprint = "1803.11167",
    archivePrefix = "arXiv",
    primaryClass = "hep-ph",
    doi = "10.1103/PhysRevD.98.075028",
    journal = "Phys. Rev. D",
    volume = "98",
    number = "7",
    pages = "075028",
    year = "2018"
}

@article{Fiorillo:2025yzf,
    author = "Fiorillo, Damiano F. G. and Pitik, Tetyana and Vitagliano, Edoardo",
    title = "{Energy Transfer by Feebly Interacting Particles in Supernovae: The Trapping Regime}",
    eprint = "2503.13653",
    archivePrefix = "arXiv",
    primaryClass = "hep-ph",
    doi = "10.1103/cz94-dqxt",
    journal = "Phys. Rev. Lett.",
    volume = "135",
    number = "7",
    pages = "071005",
    year = "2025"
}

@article{Page:2020gsx,
    author = "Page, Dany and Beznogov, Mikhail V. and Garibay, Iv{\'a}n and Lattimer, James M. and Prakash, Madappa and Janka, Hans-Thomas",
    title = "{NS 1987A in SN 1987A}",
    eprint = "2004.06078",
    archivePrefix = "arXiv",
    primaryClass = "astro-ph.HE",
    doi = "10.3847/1538-4357/ab93c2",
    journal = "Astrophys. J.",
    volume = "898",
    number = "2",
    pages = "125",
    year = "2020"
}

@article{Muller:2023pip,
    author = {M{\"u}ller, Eike and Carenza, Pierluca and Eckner, Christopher and Goobar, Ariel},
    title = "{Constraining MeV-scale axionlike particles with Fermi-LAT observations of SN 2023ixf}",
    eprint = "2306.16397",
    archivePrefix = "arXiv",
    primaryClass = "astro-ph.HE",
    reportNumber = "LAPTH-038/23",
    doi = "10.1103/PhysRevD.109.023018",
    journal = "Phys. Rev. D",
    volume = "109",
    number = "2",
    pages = "023018",
    year = "2024"
}

@article{Lee:1956qn,
    author = "Lee, T. D. and Yang, Chen-Ning",
    title = "{Question of Parity Conservation in Weak Interactions}",
    doi = "10.1103/PhysRev.104.254",
    journal = "Phys. Rev.",
    volume = "104",
    pages = "254--258",
    year = "1956"
}

@article{Wu:1957my,
    author = "Wu, C. S. and Ambler, E. and Hayward, R. W. and Hoppes, D. D. and Hudson, R. P.",
    title = "{Experimental Test of Parity Conservation in $\beta$ Decay}",
    doi = "10.1103/PhysRev.105.1413",
    journal = "Phys. Rev.",
    volume = "105",
    pages = "1413--1414",
    year = "1957"
}

@article{Pati:1974yy,
    author = "Pati, Jogesh C. and Salam, Abdus",
    title = "{Lepton Number as the Fourth Color}",
    reportNumber = "IC-74-7",
    doi = "10.1103/PhysRevD.10.275",
    journal = "Phys. Rev. D",
    volume = "10",
    pages = "275--289",
    year = "1974",
    note = "[Erratum: Phys.Rev.D 11, 703--703 (1975)]"
}

@article{Mohapatra:1974gc,
    author = "Mohapatra, R. N. and Pati, Jogesh C.",
    title = "{A Natural Left-Right Symmetry}",
    reportNumber = "CCNY-HEP-74-2",
    doi = "10.1103/PhysRevD.11.2558",
    journal = "Phys. Rev. D",
    volume = "11",
    pages = "2558",
    year = "1975"
}

@article{Senjanovic:1975rk,
    author = "Senjanovic, G. and Mohapatra, Rabindra N.",
    title = "{Exact Left-Right Symmetry and Spontaneous Violation of Parity}",
    reportNumber = "CCNY-HEP-75-5",
    doi = "10.1103/PhysRevD.12.1502",
    journal = "Phys. Rev. D",
    volume = "12",
    pages = "1502",
    year = "1975"
}

@article{Minkowski:1977sc,
    author = "Minkowski, Peter",
    title = "{$\mu \to e\gamma$ at a Rate of One Out of $10^{9}$ Muon Decays?}",
    reportNumber = "Print-77-0182 (BERN)",
    doi = "10.1016/0370-2693(77)90435-X",
    journal = "Phys. Lett. B",
    volume = "67",
    pages = "421--428",
    year = "1977"
}

@article{Mohapatra:1979ia,
    author = "Mohapatra, Rabindra N. and Senjanovic, Goran",
    title = "{Neutrino Mass and Spontaneous Parity Nonconservation}",
    reportNumber = "MDDP-TR-80-060, MDDP-PP-80-105, CCNY-HEP-79-10",
    doi = "10.1103/PhysRevLett.44.912",
    journal = "Phys. Rev. Lett.",
    volume = "44",
    pages = "912",
    year = "1980"
}

@article{Yanagida:1979as,
    author = "Yanagida, Tsutomu",
    editor = "Sawada, Osamu and Sugamoto, Akio",
    title = "{Horizontal gauge symmetry and masses of neutrinos}",
    reportNumber = "KEK-79-18-95",
    journal = "Conf. Proc. C",
    volume = "7902131",
    pages = "95--99",
    year = "1979"
}

@article{Gell-Mann:1979vob,
    author = "Gell-Mann, Murray and Ramond, Pierre and Slansky, Richard",
    title = "{Complex Spinors and Unified Theories}",
    eprint = "1306.4669",
    archivePrefix = "arXiv",
    primaryClass = "hep-th",
    reportNumber = "PRINT-80-0576",
    journal = "Conf. Proc. C",
    volume = "790927",
    pages = "315--321",
    year = "1979"
}

@article{Glashow:1979nm,
    author = "Glashow, S. L.",
    editor = "L{\'e}vy, Maurice and Basdevant, Jean-Louis and Speiser, David and Weyers, Jacques and Gastmans, Raymond and Jacob, Maurice",
    title = "{The Future of Elementary Particle Physics}",
    reportNumber = "HUTP-79-A059",
    doi = "10.1007/978-1-4684-7197-7_15",
    journal = "NATO Sci. Ser. B",
    volume = "61",
    pages = "687",
    year = "1980"
}

@article{Deshpande:1990ip,
    author = "Deshpande, N. G. and Gunion, J. F. and Kayser, Boris and Olness, Fredrick I.",
    title = "{Left-Right Symmetric Electroweak Models with Triplet Higgs}",
    reportNumber = "NSF-ITP-90-69, OITS-425",
    doi = "10.1103/PhysRevD.44.837",
    journal = "Phys. Rev. D",
    volume = "44",
    pages = "837--858",
    year = "1991"
}

@article{Maiezza:2015lza,
    author = "Maiezza, Alessio and Nemev{\v{s}}ek, Miha and Nesti, Fabrizio",
    title = "{Lepton Number Violation in Higgs Decay at LHC}",
    eprint = "1503.06834",
    archivePrefix = "arXiv",
    primaryClass = "hep-ph",
    doi = "10.1103/PhysRevLett.115.081802",
    journal = "Phys. Rev. Lett.",
    volume = "115",
    pages = "081802",
    year = "2015"
}

@article{BhupalDev:2018tox,
    author = "Bhupal Dev, P. S. and Zhang, Yongchao",
    title = "{Displaced vertex signatures of doubly charged scalars in the type-II seesaw and its left-right extensions}",
    eprint = "1808.00943",
    archivePrefix = "arXiv",
    primaryClass = "hep-ph",
    doi = "10.1007/JHEP10(2018)199",
    journal = "JHEP",
    volume = "10",
    pages = "199",
    year = "2018"
}

@article{Bambhaniya:2015ipg,
    author = "Bambhaniya, Gulab and Dev, P. S. Bhupal and Goswami, Srubabati and Mitra, Manimala",
    title = "{The Scalar Triplet Contribution to Lepton Flavour Violation and Neutrinoless Double Beta Decay in Left-Right Symmetric Model}",
    eprint = "1512.00440",
    archivePrefix = "arXiv",
    primaryClass = "hep-ph",
    reportNumber = "TUM-HEP-1028-15",
    doi = "10.1007/JHEP04(2016)046",
    journal = "JHEP",
    volume = "04",
    pages = "046",
    year = "2016"
}

@article{BhupalDev:2018xya,
    author = "Bhupal Dev, P. S. and Mohapatra, Rabindra N. and Rodejohann, Werner and Xu, Xun-Jie",
    title = "{Vacuum structure of the left-right symmetric model}",
    eprint = "1811.06869",
    archivePrefix = "arXiv",
    primaryClass = "hep-ph",
    doi = "10.1007/JHEP02(2019)154",
    journal = "JHEP",
    volume = "02",
    pages = "154",
    year = "2019"
}

@article{Chauhan:2019fji,
    author = "Chauhan, Garv",
    title = "{Vacuum Stability and Symmetry Breaking in Left-Right Symmetric Model}",
    eprint = "1907.07153",
    archivePrefix = "arXiv",
    primaryClass = "hep-ph",
    doi = "10.1007/JHEP12(2019)137",
    journal = "JHEP",
    volume = "12",
    pages = "137",
    year = "2019"
}

@article{Maiezza:2016bzp,
    author = "Maiezza, Alessio and Nemev{\v{s}}ek, Miha and Nesti, Fabrizio",
    title = "{Perturbativity and mass scales in the minimal left-right symmetric model}",
    eprint = "1603.00360",
    archivePrefix = "arXiv",
    primaryClass = "hep-ph",
    doi = "10.1103/PhysRevD.94.035008",
    journal = "Phys. Rev. D",
    volume = "94",
    number = "3",
    pages = "035008",
    year = "2016"
}

@article{Li:2020eun,
    author = "Li, Mingqiu and Yan, Qi-Shu and Zhang, Yongchao and Zhao, Zhijie",
    title = "{Prospects of gravitational waves in the minimal left-right symmetric model}",
    eprint = "2012.13686",
    archivePrefix = "arXiv",
    primaryClass = "hep-ph",
    doi = "10.1007/JHEP03(2021)267",
    journal = "JHEP",
    volume = "03",
    pages = "267",
    year = "2021"
}

@article{Searle:2025cnj,
    author = "Searle, William and Bal{\'a}zs, Csaba and Xiao, Yang and Zhang, Yang",
    title = "{Machine learning left-right breaking from gravitational waves}",
    eprint = "2506.09319",
    archivePrefix = "arXiv",
    primaryClass = "hep-ph",
    doi = "10.1088/1475-7516/2025/11/034",
    journal = "JCAP",
    volume = "11",
    pages = "034",
    year = "2025"
}

@article{Wang:2024wcs,
    author = "Wang, Dian-Wei and Yan, Qi-Shu and Huang, Mei",
    title = "{Bubble wall velocity and gravitational wave in the minimal left-right symmetric model}",
    eprint = "2405.01949",
    archivePrefix = "arXiv",
    primaryClass = "gr-qc",
    doi = "10.1103/PhysRevD.110.076011",
    journal = "Phys. Rev. D",
    volume = "110",
    number = "7",
    pages = "076011",
    year = "2024"
}

@article{Brdar:2019fur,
    author = "Brdar, Vedran and Graf, Lukas and Helmboldt, Alexander J. and Xu, Xun-Jie",
    title = "{Gravitational Waves as a Probe of Left-Right Symmetry Breaking}",
    eprint = "1909.02018",
    archivePrefix = "arXiv",
    primaryClass = "hep-ph",
    doi = "10.1088/1475-7516/2019/12/027",
    journal = "JCAP",
    volume = "12",
    pages = "027",
    year = "2019"
}

@article{Jiang:2025nie,
    author = "Jiang, Xu-Hui and Lu, Chih-Ting",
    title = "{Searching for axion-like particles from tau exotic decays at the Super Tau-Charm Facility and its far detectors}",
    eprint = "2509.23949",
    archivePrefix = "arXiv",
    primaryClass = "hep-ph",
    reportNumber = "CPTNP-2025-038",
    month = "9",
    year = "2025"
}

@article{Borboruah:2022eex,
    author = "Borboruah, Z. A. and Yajnik, U. A.",
    title = "{Left-right symmetry breaking and gravitational waves: A tale of two phase transitions}",
    eprint = "2212.05829",
    archivePrefix = "arXiv",
    primaryClass = "astro-ph.CO",
    doi = "10.1103/PhysRevD.110.043016",
    journal = "Phys. Rev. D",
    volume = "110",
    number = "4",
    pages = "043016",
    year = "2024"
}

@article{Borah:2022wdy,
    author = "Borah, Debasish and Dasgupta, Arnab",
    title = "{Probing left-right symmetry via gravitational waves from domain walls}",
    eprint = "2205.12220",
    archivePrefix = "arXiv",
    primaryClass = "hep-ph",
    doi = "10.1103/PhysRevD.106.035016",
    journal = "Phys. Rev. D",
    volume = "106",
    number = "3",
    pages = "035016",
    year = "2022"
}

@article{Fukugita:1986hr,
    author = "Fukugita, M. and Yanagida, T.",
    title = "{Baryogenesis Without Grand Unification}",
    reportNumber = "RIFP-641",
    doi = "10.1016/0370-2693(86)91126-3",
    journal = "Phys. Lett. B",
    volume = "174",
    pages = "45--47",
    year = "1986"
}

@article{Frere:2008ct,
    author = "Frere, Jean-Marie and Hambye, Thomas and Vertongen, Gilles",
    title = "{Is leptogenesis falsifiable at LHC?}",
    eprint = "0806.0841",
    archivePrefix = "arXiv",
    primaryClass = "hep-ph",
    reportNumber = "ULB-TH-08-17, NSF-KITP-08-79",
    doi = "10.1088/1126-6708/2009/01/051",
    journal = "JHEP",
    volume = "01",
    pages = "051",
    year = "2009"
}

@article{ReFiorentin:2016rzn,
    author = "Re Fiorentin, M. and Niro, V. and Fornengo, N.",
    title = "{A consistent model for leptogenesis, dark matter and the IceCube signal}",
    eprint = "1606.04445",
    archivePrefix = "arXiv",
    primaryClass = "hep-ph",
    reportNumber = "FTUAM-16-22, IFT-UAM-CSIC-16-055",
    doi = "10.1007/JHEP11(2016)022",
    journal = "JHEP",
    volume = "11",
    pages = "022",
    year = "2016"
}

@article{Hufnagel:2017dgo,
    author = "Hufnagel, Marco and Schmidt-Hoberg, Kai and Wild, Sebastian",
    title = "{BBN constraints on MeV-scale dark sectors. Part I. Sterile decays}",
    eprint = "1712.03972",
    archivePrefix = "arXiv",
    primaryClass = "hep-ph",
    reportNumber = "DESY-17-211",
    doi = "10.1088/1475-7516/2018/02/044",
    journal = "JCAP",
    volume = "02",
    pages = "044",
    year = "2018"
}

@article{JUNO:2025gmd,
    author = "Abusleme, Angel and others",
    collaboration = "JUNO",
    title = "{First measurement of reactor neutrino oscillations at JUNO}",
    eprint = "2511.14593",
    archivePrefix = "arXiv",
    primaryClass = "hep-ex",
    month = "11",
    year = "2025"
}

@article{Depta:2020zbh,
    author = "Depta, Paul Frederik and Hufnagel, Marco and Schmidt-Hoberg, Kai",
    title = "{Updated BBN constraints on electromagnetic decays of MeV-scale particles}",
    eprint = "2011.06519",
    archivePrefix = "arXiv",
    primaryClass = "hep-ph",
    reportNumber = "DESY-20-160, DESY 20-160, ULB-TH/20-15",
    doi = "10.1088/1475-7516/2021/04/011",
    journal = "JCAP",
    volume = "04",
    pages = "011",
    year = "2021"
}

@article{Langhoff:2022bij,
    author = "Langhoff, Kevin and Outmezguine, Nadav Joseph and Rodd, Nicholas L.",
    title = "{Irreducible Axion Background}",
    eprint = "2209.06216",
    archivePrefix = "arXiv",
    primaryClass = "hep-ph",
    reportNumber = "CERN-TH-2022-148",
    doi = "10.1103/PhysRevLett.129.241101",
    journal = "Phys. Rev. Lett.",
    volume = "129",
    number = "24",
    pages = "241101",
    year = "2022"
}

@article{Cadamuro:2011fd,
    author = "Cadamuro, Davide and Redondo, Javier",
    title = "{Cosmological bounds on pseudo Nambu-Goldstone bosons}",
    eprint = "1110.2895",
    archivePrefix = "arXiv",
    primaryClass = "hep-ph",
    reportNumber = "MPP-2011-116",
    doi = "10.1088/1475-7516/2012/02/032",
    journal = "JCAP",
    volume = "02",
    pages = "032",
    year = "2012"
}

@article{Depta:2020wmr,
    author = "Depta, Paul Frederik and Hufnagel, Marco and Schmidt-Hoberg, Kai",
    title = "{Robust cosmological constraints on axion-like particles}",
    eprint = "2002.08370",
    archivePrefix = "arXiv",
    primaryClass = "hep-ph",
    reportNumber = "DESY-20-003, DESY 20-003",
    doi = "10.1088/1475-7516/2020/05/009",
    journal = "JCAP",
    volume = "05",
    pages = "009",
    year = "2020"
}

@article{Balazs:2022tjl,
    author = "Bal{\'a}zs, Csaba and others",
    title = "{Cosmological constraints on decaying axion-like particles: a global analysis}",
    eprint = "2205.13549",
    archivePrefix = "arXiv",
    primaryClass = "astro-ph.CO",
    reportNumber = "gambit-physics-2022, KCL-PH-TH/2022-23, TTP22-034",
    doi = "10.1088/1475-7516/2022/12/027",
    journal = "JCAP",
    volume = "12",
    pages = "027",
    year = "2022"
}

@article{Berger:2016vxi,
    author = "Berger, Joshua and Jedamzik, Karsten and Walker, Devin G. E.",
    title = "{Cosmological Constraints on Decoupled Dark Photons and Dark Higgs}",
    eprint = "1605.07195",
    archivePrefix = "arXiv",
    primaryClass = "hep-ph",
    reportNumber = "SLAC-PUB-16533",
    doi = "10.1088/1475-7516/2016/11/032",
    journal = "JCAP",
    volume = "11",
    pages = "032",
    year = "2016"
}

@article{Fradette:2018hhl,
    author = "Fradette, Anthony and Pospelov, Maxim and Pradler, Josef and Ritz, Adam",
    title = "{Cosmological beam dump: constraints on dark scalars mixed with the Higgs boson}",
    eprint = "1812.07585",
    archivePrefix = "arXiv",
    primaryClass = "hep-ph",
    doi = "10.1103/PhysRevD.99.075004",
    journal = "Phys. Rev. D",
    volume = "99",
    number = "7",
    pages = "075004",
    year = "2019"
}

@article{DEramo:2024lsk,
    author = "D'Eramo, Francesco and Tesi, Andrea and Vaskonen, Ville",
    title = "{Irreducible cosmological backgrounds of a real scalar with a broken symmetry}",
    eprint = "2407.19997",
    archivePrefix = "arXiv",
    primaryClass = "hep-ph",
    doi = "10.1103/PhysRevD.110.095002",
    journal = "Phys. Rev. D",
    volume = "110",
    number = "9",
    pages = "095002",
    year = "2024"
}

@article{Akita:2024nam,
    author = "Akita, Kensuke and Baur, Gideon and Ovchynnikov, Maksym and Schwetz, Thomas and Syvolap, Vsevolod",
    title = "{New Physics Decaying into Metastable Particles: Impact on Cosmic Neutrinos}",
    eprint = "2411.00892",
    archivePrefix = "arXiv",
    primaryClass = "hep-ph",
    reportNumber = "CERN-TH-2024-188, CERN-TH-2024-188",
    doi = "10.1103/PhysRevLett.134.121001",
    journal = "Phys. Rev. Lett.",
    volume = "134",
    number = "12",
    pages = "121001",
    year = "2025"
}

@article{Jung:2025dyo,
    author = "Jung, Tae Hyun and Okui, Takemichi and Tobioka, Kohsaku and Wang, Jiabao",
    title = "{New bounds on heavy QCD axions from big bang nucleosynthesis}",
    eprint = "2510.23695",
    archivePrefix = "arXiv",
    primaryClass = "hep-ph",
    reportNumber = "CTPU-PTC-25-30",
    doi = "10.1103/l2m1-h1cp",
    journal = "Phys. Rev. D",
    volume = "113",
    number = "5",
    pages = "055002",
    year = "2026"
}

@article{Chang:1983fu,
    author = "Chang, D. and Mohapatra, R. N. and Parida, M. K.",
    title = "{Decoupling Parity and $SU(2)_R$ Breaking Scales: A New Approach to Left-Right Symmetric Models}",
    reportNumber = "MdDP-TR-84-65",
    doi = "10.1103/PhysRevLett.52.1072",
    journal = "Phys. Rev. Lett.",
    volume = "52",
    pages = "1072",
    year = "1984"
}

@article{Cirigliano:2004tc,
    author = "Cirigliano, V. and Kurylov, A. and Ramsey-Musolf, M. J. and Vogel, P.",
    title = "{Neutrinoless double beta decay and lepton flavor violation}",
    eprint = "hep-ph/0406199",
    archivePrefix = "arXiv",
    doi = "10.1103/PhysRevLett.93.231802",
    journal = "Phys. Rev. Lett.",
    volume = "93",
    pages = "231802",
    year = "2004"
}

@article{Nemevsek:2011aa,
    author = "Nemevsek, Miha and Nesti, Fabrizio and Senjanovic, Goran and Tello, Vladimir",
    title = "{Neutrinoless Double Beta Decay: Low Left-Right Symmetry Scale?}",
    eprint = "1112.3061",
    archivePrefix = "arXiv",
    primaryClass = "hep-ph",
    month = "12",
    year = "2011"
}

@article{Alves:2022yav,
    author = "Alves, Gustavo F. S. and Fong, Chee Sheng and Leal, Luighi P. S. and Funchal, Renata Zukanovich",
    title = "{Exploring the Neutrino Sector of the Minimal Left-Right Symmetric Model}",
    eprint = "2208.07378",
    archivePrefix = "arXiv",
    primaryClass = "hep-ph",
    month = "8",
    year = "2022"
}

@article{Borah:2016iqd,
    author = "Borah, Debasish and Dasgupta, Arnab",
    title = "{Charged lepton flavour violcxmation and neutrinoless double beta decay in left-right symmetric models with type I+II seesaw}",
    eprint = "1606.00378",
    archivePrefix = "arXiv",
    primaryClass = "hep-ph",
    doi = "10.1007/JHEP07(2016)022",
    journal = "JHEP",
    volume = "07",
    pages = "022",
    year = "2016"
}

@article{Deppisch:2014zta,
    author = "Deppisch, Frank F. and Gonzalo, Tomas E. and Patra, Sudhanwa and Sahu, Narendra and Sarkar, Utpal",
    title = "{Double beta decay, lepton flavor violation, and collider signatures of left-right symmetric models with spontaneous $D$-parity breaking}",
    eprint = "1410.6427",
    archivePrefix = "arXiv",
    primaryClass = "hep-ph",
    reportNumber = "LCTS-2014-42",
    doi = "10.1103/PhysRevD.91.015018",
    journal = "Phys. Rev. D",
    volume = "91",
    number = "1",
    pages = "015018",
    year = "2015"
}

@article{Barry:2013xxa,
    author = "Barry, James and Rodejohann, Werner",
    title = "{Lepton number and flavour violation in TeV-scale left-right symmetric theories with large left-right mixing}",
    eprint = "1303.6324",
    archivePrefix = "arXiv",
    primaryClass = "hep-ph",
    doi = "10.1007/JHEP09(2013)153",
    journal = "JHEP",
    volume = "09",
    pages = "153",
    year = "2013"
}

@article{Lee:2013htl,
    author = "Lee, Chang-Hun and Bhupal Dev, P. S. and Mohapatra, R. N.",
    title = "{Natural TeV-scale left-right seesaw mechanism for neutrinos and experimental tests}",
    eprint = "1309.0774",
    archivePrefix = "arXiv",
    primaryClass = "hep-ph",
    reportNumber = "UMD-PP-013-011, MAN-HEP-2013-21",
    doi = "10.1103/PhysRevD.88.093010",
    journal = "Phys. Rev. D",
    volume = "88",
    number = "9",
    pages = "093010",
    year = "2013"
}

@article{Awasthi:2013ff,
    author = "Awasthi, Ram Lal and Parida, M. K. and Patra, Sudhanwa",
    title = "{Neutrino masses, dominant neutrinoless double beta decay, and observable lepton flavor violation in left-right models and SO(10) grand unification with low mass $ W_R, Z_R$ bosons}",
    eprint = "1302.0672",
    archivePrefix = "arXiv",
    primaryClass = "hep-ph",
    doi = "10.1007/JHEP08(2013)122",
    journal = "JHEP",
    volume = "08",
    pages = "122",
    year = "2013"
}

@article{Qiang:2026abc,
    author = "Qiang, Shufang and Wu, Peiwen and Zhang, Yongchao",
    title = "{1-loop phenomenology of light $SU(2)_R$-breaking scalar in the minimal left-right symmetric model}",
    note = {in progress},
    year = "2026"
}

@article{Bryman:1986wn,
    author = "Bryman, D. A. and Clifford, E. T. H.",
    title = "{EXOTIC MUON DECAY $\mu \rightarrow e + x$}",
    reportNumber = "TRI-PP-86-73",
    doi = "10.1103/PhysRevLett.57.2787",
    journal = "Phys. Rev. Lett.",
    volume = "57",
    pages = "2787",
    year = "1986"
}

@article{Hahn:1998yk,
    author = "Hahn, T. and Perez-Victoria, M.",
    title = "{Automatized one loop calculations in four-dimensions and D-dimensions}",
    eprint = "hep-ph/9807565",
    archivePrefix = "arXiv",
    reportNumber = "UG-FT-87-98, KA-TP-7-1998",
    doi = "10.1016/S0010-4655(98)00173-8",
    journal = "Comput. Phys. Commun.",
    volume = "118",
    pages = "153--165",
    year = "1999"
}

@article{Khan:2019doq,
    author = "Khan, Amir N. and Nunokawa, Hiroshi and Parke, Stephen J",
    title = "{Why matter effects matter for JUNO}",
    eprint = "1910.12900",
    archivePrefix = "arXiv",
    primaryClass = "hep-ph",
    reportNumber = "FERMILAB-PUB-19-490-T",
    doi = "10.1016/j.physletb.2020.135354",
    journal = "Phys. Lett. B",
    volume = "803",
    pages = "135354",
    year = "2020"
}

@article{Candon:2025ypl,
    author = "Cand{\'o}n, Francisco R. and Fiorillo, Damiano F. G. and Janka, Hans-Thomas and van Baal, Bart F. A. and Vitagliano, Edoardo",
    title = "{Small Progenitors, Large Couplings: Type Ic Supernova Constraints on Radiatively Decaying Particles}",
    eprint = "2509.18253",
    archivePrefix = "arXiv",
    primaryClass = "hep-ph",
    month = "9",
    year = "2025"
}

@misc{data,
  author = {Qiang, Shufang and Wu, Peiwen and Zhang, Yongchao},
  howpublished = "\url{https://pan.seu.edu.cn/link/AA5AD8624CF35549CF9894A11C715FC103}",
  year = {2026}
}

@article{Patrone:2025fwk,
    author = "Patrone, Samuel and Blinov, Nikita and Plestid, Ryan",
    title = "{Long-lived axionlike particles from electromagnetic cascades}",
    eprint = "2509.14310",
    archivePrefix = "arXiv",
    primaryClass = "hep-ph",
    reportNumber = "CALT-TH/2025-029, CERN-TH-2025-178",
    doi = "10.1103/38hq-qk7k",
    journal = "Phys. Rev. D",
    volume = "113",
    number = "7",
    pages = "075009",
    year = "2026"
}

@article{Bao:2025tqs,
    author = "Bao, Shou-shan and Ma, Yang and Wu, Yongcheng and Xie, Keping and Zhang, Hong",
    title = "{Light axion-like particles at future lepton colliders}",
    eprint = "2505.10023",
    archivePrefix = "arXiv",
    primaryClass = "hep-ph",
    reportNumber = "COMETA-2025-02, IRMP-CP3-25-09, MSUHEP-25-002, CPTNP-2025-011",
    doi = "10.1007/JHEP10(2025)122",
    journal = "JHEP",
    volume = "10",
    pages = "122",
    year = "2025"
}
\bibliographystyle{JHEP}

\end{document}